\documentclass[11pt,a4paper]{article}
\usepackage{jheppub}
\usepackage{comment}
\pdfoutput=1
\usepackage{dcolumn}% Align table columns on decimal point
\usepackage{bm}% bold math
\usepackage{graphicx}% Include figure files
\usepackage{diagbox}
\usepackage{amssymb,amsmath}
\usepackage{multirow}
\usepackage{color,url}
\usepackage{tabu}
\usepackage{array}
\usepackage[normalem]{ulem}
\usepackage{soul}
\usepackage{colordvi}
\usepackage{subcaption}
\usepackage{mathrsfs}
\usepackage{enumitem}
\def\ga{\mathrel{\raise.3ex\hbox{$$>$$\kern-.75em\lower1ex\hbox{$\sim$}}}}
\def\la{\mathrel{\raise.3ex\hbox{$<$\kern-.75em\lower1ex\hbox{$\sim$}}}}
\def\beqa{\begin{eqnarray}}
\def\eeqa{\end{eqnarray}}
\usepackage{slashed}

\title{Measuring the trilinear Higgs self-coupling in Higgs boson pair production at multi-TeV muon colliders}

\author[a,b]{Kingman Cheung,}
\author[c]{Jae Sik Lee,}
\author[a,b]{Soojin Lee,}
\author[a,b]{Chen Wang}

\affiliation[a]{Department of Physics, National Tsing Hua University, Hsinchu 300, Taiwan}
\affiliation[b]{Center for Theory and Computation, National Tsing Hua University, Hsinchu 300, Taiwan}
\affiliation[c]{Department of Physics and IUEP, Chonnam National University, Gwangju 61186, South Korea}

\emailAdd{cheung@phys.nthu.edu.tw}
\emailAdd{jslee@jnu.ac.kr}
\emailAdd{soojin.lee@gapp.nthu.edu.tw}
\emailAdd{elian890415383@gmail.com}

\abstract{
The trilinear Higgs self-coupling determines the shape of the Higgs potential, and its measurement is a central goal of future colliders. We assess the sensitivity of multi-TeV muon colliders to the coupling modifier $\kappa_3$ in Higgs boson pair production via vector boson fusion, using the $b\bar{b}b\bar{b}$ final state at $\sqrt{s}=3$ TeV with $1$ ab$^{-1}$ and at $10$ TeV with $10$ ab$^{-1}$. Events are analyzed in two complementary regions, a resolved region with four jets and a boosted region with two large-radius jets. To extract the signal from backgrounds a few orders of magnitude larger, we combine a supervised jet-to-Higgs pairing network based on the SPANet approach, topological data analysis of the event energy flow, and two dedicated classifiers, $\mathcal{D}_{\rm HH}$ for the signal-to-background separation and $\mathcal{D}_{\kappa_3}$ for the $\kappa_3$ shape information. The coupling is extracted from a two-dimensional likelihood fit to the distribution of the two classifier outputs. Combining the two regions, we obtain $0.80<\kappa_3<1.29$ at $3$ TeV and $0.96<\kappa_3<1.05$ at $10$ TeV at $68\%$ confidence level. The $10$ TeV determination reaches the few-percent level in this statistics-limited projection, which surpasses by a large amount the precision projected for the HL-LHC.
}

          
\begin{document}
\maketitle

%--------------------------------------------------------------------------------
\section{Introduction}\label{sec:introduction}

The origin of mass is one of the outstanding questions in modern particle physics, which is related to the mechanism involved in electroweak symmetry breaking (EWSB), which is believed to give mass to matter and gauge bosons. The simplest implementation of EWSB in the standard model (SM) is to introduce a Higgs doublet field~\cite{Higgs:1964pj,Englert:1964et,Guralnik:1964eu}. With the discovery of an SM-like Higgs boson in 2012~\cite{ATLAS:2012yve,CMS:2012qbp} and all the follow-up measurements of its properties, the identity of the SM Higgs boson has been established, with only small room for deviation.

However, there is no {\it a priori} reason why the EWSB sector simply contains only one Higgs doublet field. Indeed, many new physics models involve extensions of the EWSB sector~\cite{Gunion:1989we,Ivanov:2017dad}. For example, supersymmetry is a well-known solution to the gauge hierarchy problem and has at least two Higgs doublet fields~\cite{Nilles:1983ge,Haber:1984rc,Martin:1997ns}. Therefore, probing the EWSB sector may give insights into ultraviolet physics at the TeV scale. Such extended scalar sectors can also be searched for directly at a multi-TeV muon collider, for instance through the pair production of additional neutral scalars~\cite{Coleppa:2026esg}.

One of the best ways to probe the structure of the EWSB sector is to probe the Higgs self-couplings. This is because the self-couplings of the Higgs boson are very different among the SM, two Higgs doublet models (2HDM)~\cite{Lee:1973iz,Branco:2011iw}, the minimal supersymmetric SM (MSSM)~\cite{Haber:1984rc,Djouadi:2005gj}, and any composite Higgs models~\cite{Kaplan:1983fs,Georgi:1984af,Panico:2015jxa}. 
The trilinear coupling is moreover of cosmological interest, since a modified Higgs potential can drive a first-order electroweak phase transition and source gravitational waves~\cite{Jahedi:2025yjz}.

Higgs boson pair ($HH$) production at the LHC serves as a principal probe of the trilinear Higgs self-coupling~\cite{Glover:1987nx,Dicus:1987ic,Plehn:1996wb,Djouadi:1999rca,Dawson:1998py,Baur:2002qd,Binoth:2006ym,Baur:2003gpa,Baglio:2012np,Grigo:2013rya,Barger:2013jfa,Adhikary:2017jtu,Adhikary:2020fqf}. $HH$ production beyond the SM (BSM), including resonant production of heavy Higgs bosons~\cite{Adhikary:2018ise}, has also been studied extensively, as reviewed for example in Ref.~\cite{Lu:2015jza} and references therein. Nevertheless, the trilinear coupling remains only loosely constrained at the LHC, $-0.71<\kappa_3<6.1$ at $95\%$ confidence level~\cite{CMS:2026nuu}, and even the projected sensitivity of the high-luminosity LHC stays at the $\mathcal{O}(50\%)$ level~\cite{ATLAS:2022faz}.

We therefore turn to multi-TeV muon colliders~\cite{Palmer:1996gs,Ankenbrandt:1999cta,Delahaye:2019omf,AlAli:2021let,Schulte:2021eyr,Black:2022cth,Accettura:2023ked}. Because the muon is point-like and loses little energy to synchrotron radiation, essentially the full collision energy is available for the hard scattering in a clean leptonic environment, and the rate of $HH$ production via vector boson fusion (VBF) grows with the center-of-mass energy~\cite{Costantini:2020stv,Han:2020pif}.
We investigate the potential of the $3$\,TeV and $10$\,TeV muon colliders in testing the trilinear Higgs coupling via $HH$ production and hadronic decay of the Higgs pair into $b\bar b b\bar b$ in order to maximize the signal yield. Nevertheless, such a signal faces a number of SM backgrounds, the total of which is a few orders of magnitude larger.

We identify two cases for our analysis. Before we lay out these two cases, we note that the SM background refers to the SM processes which mimic the $4b$ final state of the signal $HH$ production shown in Fig.~\ref{fig:feynman_diagrams}, and that it excludes $HH$ production itself. The exclusion holds for every value of $\kappa_3$, because the two amplitudes without the trilinear vertex keep $HH$ production finite even at $\kappa_3=0$, so $HH$ events always belong to the signal and never to the background. Here $\kappa_3$ denotes the trilinear coupling modifier defined in Sec.~\ref{sec:sm_processes}, and in both cases it is scanned over the range $0.2 \le \kappa_3 \le 1.8$.

\begin{enumerate}[label=\textbf{Case~\arabic*:}, leftmargin=*, align=left]
\item Observability of $HH$ production above the SM background.
  \begin{flalign*}
    \quad & \begin{array}{@{}l@{\;=\;}l@{}}
      \text{null hypothesis} & \text{SM background}\\
      \text{test hypothesis} & \text{SM background} \,+\, HH \text{ production}
    \end{array} &
  \end{flalign*}

\item Discrimination of $HH$ production with $\kappa_3 \neq 1$ from that with $\kappa_3=1$.
  \begin{flalign*}
    \quad & \begin{array}{@{}l@{\;=\;}l@{}}
      \text{null hypothesis} & \text{SM background} \,+\, HH \text{ production with } \kappa_3 = 1\\
      \text{test hypothesis} & \text{SM background} \,+\, HH \text{ production with } \kappa_3 \neq 1
    \end{array} &
  \end{flalign*}
\end{enumerate}

To address these two cases in the presence of large backgrounds, our analysis rests on the following three key strategies.

\begin{enumerate}
\item {\bf Resolved and boosted regions in simulations.}
  The final state appears differently in two kinematic regions separated by the Higgs transverse momentum. At $p_T < 200$ GeV the Higgs boson decays into a pair of $b\bar b$ with a wider angular separation $\Delta R_{b\bar b} \gtrsim 1$, so the two $b$ quarks can be identified separately. We refer to this as the resolved region.
  On the other hand, at high transverse momentum $p_T$, the Higgs boson decays into a collimated pair of $b\bar b$, which appears as a single fat Higgs jet. We refer to this as the boosted region. At the end, we combine the likelihood of both regions to improve the detectability.

\item {\bf Inclusion of TDA variables in the machine learning (ML) 
models.}
  Detecting $HH$ production in the $4b$ mode is difficult. In addition to the conventional kinematic variables, we introduce topological data analysis (TDA) descriptors~\cite{Edelsbrunner:2002}, which summarize the global $(\eta,\phi)$ energy flow topology of each event through persistent homology and provide information complementary to the standard observables. Such topological methods have only recently entered collider phenomenology~\cite{Li:2020jdb,Beuria:2023seb,Beuria:2025avt,Mittal:2026tss}.

\item {\bf Two-network design.}
  We adopt a two-network machine learning strategy inspired by Refs.~\cite{Andreetto:2024rra,Andreetto:2026flp,Andreetto:2026gvm,Andreetto:2026pyu}. One network, $\mathcal{D}_{\rm HH}$, is trained to separate the $HH$ signal from the SM background (Case~1), and the other, $\mathcal{D}_{\kappa_3}$, to discriminate among trilinear coupling hypotheses (Case~2). Both are applied to each event, and their outputs are combined in a two-dimensional likelihood fit from which $\kappa_3$ is extracted. In the resolved region the four jets are first assigned to the two Higgs candidates by a dedicated network based on the Symmetry Preserving Attention Network (SPANet) approach~\cite{Shmakov:2021qdz}.
\end{enumerate}

The paper is organized as follows. In Sec.~\ref{sec:sm_processes}, we set up the Higgs sector and notation, and identify the signal and background processes together with their generation.
In Sec.~\ref{sec:event_selection}, we describe the event reconstruction and the pre-selection of the resolved and boosted regions. In Sec.~\ref{sec:ml_likelihood}, we introduce the TDA descriptors, describe the machine learning models and interpret their input features, and set up the two-dimensional likelihood analysis used to extract $\kappa_3$.
In Sec.~\ref{sec:results}, we present the resulting sensitivities for both the $3$\,TeV and $10$\,TeV muon colliders. We conclude in Sec.~\ref{sec:conclusion}.

%--------------------------------------------------------------------------------
\section{Higgs Boson Pair Production and Event Generation}\label{sec:sm_processes}

\subsection{Higgs boson pair production in the Standard Model}\label{sec:hh_production}

The Higgs sector of the SM~\cite{Glashow:1961tr, Weinberg:1967tq, Salam:1968rm} is fully encoded in the scalar potential~\cite{Englert:1964et,Higgs:1964pj,Guralnik:1964eu}
\begin{equation} \label{eq:potential}
V(H) \;=\; -\mu^2\,H^\dagger H \;+\; \lambda\,(H^\dagger H)^2,
\end{equation}
which, after electroweak symmetry breaking, gives rise to a single physical Higgs boson with mass $m_H = 125$\,GeV~\cite{ATLAS:2012yve, CMS:2012qbp} and to a trilinear self-coupling $\lambda_3^{\rm SM} = m_H^2/(2v) = \lambda\,v$, where $v = 246$\,GeV is the electroweak vacuum expectation value. 
Because $\lambda_3$ controls the curvature of the potential at its minimum, any deviation from the SM value would signal physics beyond the SM in the Higgs sector and a non-trivial shape of the electroweak vacuum. It is conventional to encode the self-coupling in the dimensionless ratio
\begin{equation}\label{eq:kappa3_def}
\kappa_3 \;\equiv\; \lambda_3/\lambda_3^{\rm SM},
\end{equation}
so that $\kappa_3=1$ corresponds to the SM. 
The current LHC determinations of the Higgs couplings relevant to $HH$ production, of which only $\kappa_3$ is varied in this work, sit at qualitatively different precision levels. Here $\kappa_W \equiv g_{HWW}/g_{HWW}^{\rm SM}$, $\kappa_Z \equiv g_{HZZ}/g_{HZZ}^{\rm SM}$, and $\kappa_{2V} \equiv g_{HHVV}/g_{HHVV}^{\rm SM}$ are the single- and double-Higgs gauge coupling modifiers, defined analogously to Eq.~(\ref{eq:kappa3_def}), and we write $\kappa_V \equiv \kappa_W = \kappa_Z$ for their common value.
\begin{itemize}
\item $\kappa_W$: $\kappa_W = 1.02 \pm 0.08$ (CMS, $\pm 1\sigma$, best-fit)~\cite{CMS:2022dwd};
\item $\kappa_Z$: $\kappa_Z = 1.04 \pm 0.07$ (CMS, $\pm 1\sigma$, best-fit)~\cite{CMS:2022dwd};
\item $\kappa_{2V}$: $0.73 < \kappa_{2V} < 1.3$ (95\% CL, ATLAS+CMS combination)~\cite{CMS:2026nuu};
\item $\kappa_3$: $-0.71 < \kappa_3 < 6.1$ (95\% CL, ATLAS+CMS combination)~\cite{CMS:2026nuu}.
\end{itemize}
$\kappa_W$ and $\kappa_Z$ are already determined at the percent level, $\kappa_{2V}$ at the $\mathcal{O}(30\%)$ level, but the trilinear $\kappa_3$ remains by far the loosest and even the projected high-luminosity LHC sensitivity stays at the $\mathcal{O}(50\%)$ level~\cite{ATLAS:2022faz}, insufficient to pin down the shape of the potential or to constrain extended Higgs sectors. In this work we therefore take $\kappa_V=\kappa_{2V}=1$ and probe the dependence on $\kappa_3$ alone, with the goal of compressing the wide $\kappa_3$ window at a multi-TeV muon collider~\cite{Palmer:1996gs,Ankenbrandt:1999cta,AlAli:2021let,Schulte:2021eyr,Black:2022cth,Accettura:2023ked}.

\begin{figure}[t]
    \centering
    \includegraphics[width=\linewidth]{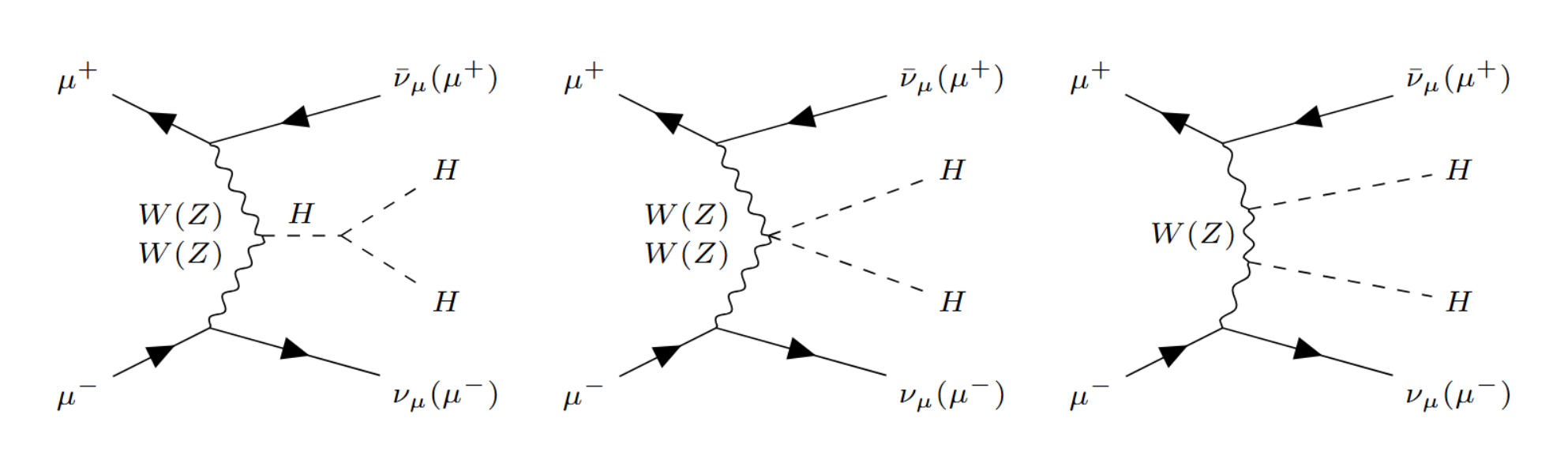}
    \caption{Feynman diagrams of Higgs boson pair production via vector boson fusion at a muon collider, $\mu^+\mu^-\to HH\,X$ with $X\in\{\nu_\mu\bar\nu_\mu,\,\mu^+\mu^-\}$.}
    \label{fig:feynman_diagrams}
\end{figure}

A multi-TeV muon collider provides a qualitatively different and clean probe of the Higgs self-coupling~\cite{Han:2020pif,Andreetto:2024rra,Andreetto:2026pyu}. As noted in Sec.~\ref{sec:introduction}, essentially the full collision energy is available for the hard scattering, and the dominant $HH$ production mechanism is VBF, whose cross section grows monotonically with $\sqrt{s}$.
At hadron colliders, by contrast, $HH$ production is dominated by loop-induced gluon fusion through top quark triangle and box diagrams, which have no counterpart at a muon collider. The remaining tree-level diagrams, which proceed through the muon Yukawa coupling as in the $s$-channel $\mu^+\mu^-\to H^\ast\to HH$, are negligible owing to its smallness. Three amplitudes contribute coherently to $\mu^+\mu^-\to HH\,X$, in both the charged current ($X=\nu_\mu\bar\nu_\mu$, $W$-fusion) and the neutral current ($X=\mu^+\mu^-$, $Z$-fusion) channels, as Fig.~\ref{fig:feynman_diagrams} shows from left to right. They are (i) the $\kappa_3$-sensitive diagram $VV \to H^\ast \to HH$ through an off-shell Higgs and the trilinear vertex, (ii) the quartic $HHVV$ diagram $VV\to HH$, and (iii) the diagrams $VV\to HH$ involving two $HVV$ vertices.

\begin{figure}[t]
\centering
\includegraphics[width=0.7\textwidth]{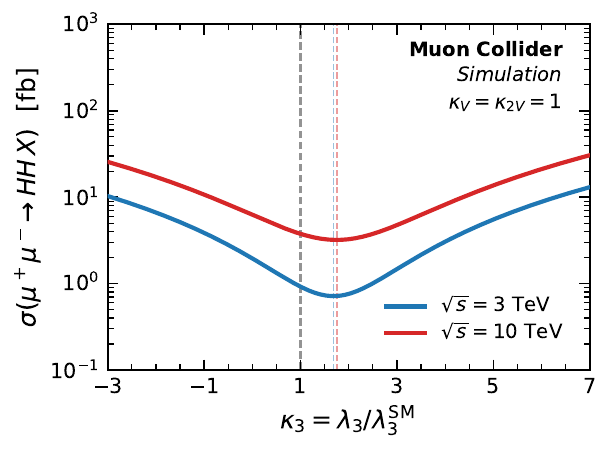}
\caption{Total production cross section $\sigma(\mu^+\mu^-\!\to HH\,X)$, {$X=\nu_\mu\bar\nu_\mu+\mu^+\mu^-$} (sum of $W$- and $Z$-fusion final states), as a function of $\kappa_3$ at $\kappa_V=\kappa_{2V}=1$, for $\sqrt{s}=3$\,TeV (blue) and $10$\,TeV (red). The vertical dashed gray line marks the SM ($\kappa_3=1$), and the vertical colored dashed lines indicate the position of the cross section minimum for each energy.}
\label{fig:xsec_kappa3}
\end{figure}

Since we hold $\kappa_V=\kappa_{2V}=1$, the cross section reduces to a quadratic function of the trilinear coupling alone with coefficients that depend on $\sqrt{s}$ and the applied selection cuts.
Fig.~\ref{fig:xsec_kappa3} shows the total cross section, the sum of the charged and neutral current channels, as a function of $\kappa_3$ at $\sqrt{s}=3$\,TeV (blue) and $10$\,TeV (red), with both muons in the neutral current final state required to satisfy $|\eta|>2.5$ and thus escape the detector acceptance. Destructive interference among the three contributing amplitudes produces a minimum at $\kappa_3^{\rm min}\!\simeq\!1.68$ at $3$\,TeV and $1.75$ at $10$\,TeV. As a consequence, $\kappa_3=1$ sits on the steep left flank of the parabola, where the cross section varies more rapidly for $\kappa_3<1$ than for $\kappa_3>1$.
This intrinsic asymmetry will be inherited by any analysis sensitive to the total rate, so the lower bound on $\kappa_3$ will be tighter than the upper one. In this work we focus on the $HH\to b\bar bb\bar b$ final state, which combines the largest branching fraction ($[\mathrm{BR}(H\to b\bar b)]^2\simeq 0.34$\footnote{The branching ratios follow the LHC Higgs Working Group recommendation, \url{https://twiki.cern.ch/twiki/bin/view/LHCPhysics/CERNYellowReportPageBR} with $m_H=125.00$~GeV.}) with manageable physics backgrounds at a muon collider.
At $\kappa_3=1$ the resulting inclusive signal cross section is $\sigma\times[\mathrm{BR}(H\to b\bar b)]^2 = 0.312$\,fb at $\sqrt{s}=3$\,TeV and $1.273$\,fb at $10$\,TeV, as also listed in Table~\ref{tab:preselection_resolved}.
Near the SM point the trilinear amplitude is subdominant in the total rate, so its distinctive enhancement of the $HH$ invariant mass near threshold is largely masked by the non-trilinear contributions. Isolating the $\kappa_3$ information therefore requires a dedicated multivariate analysis of the phase-space regions where the trilinear amplitude leaves a visible footprint, as developed in Sec.~\ref{sec:ml_likelihood}.

\subsection{Signal and background processes}\label{sec:signal_background}

Following the standard LHC strategy~\cite{CMS:2026nuu}, we divide the $HH\to b\bar{b}b\bar{b}$ final state into two kinematic regions according to the transverse momentum of the Higgs candidates, a resolved region where $p_T(H)<200$\,GeV and a boosted region where $p_T(H)>200$\,GeV, roughly the momentum above which the two $b$ quarks from $H\to b\bar b$, separated by $\Delta R_{b\bar b}\approx 2m_H/p_T(H)$, begin to be captured by a single $R=1.0$ jet.

At the truth level, $77.7\%$ of SM signal events ($\kappa_3=1$) at $\sqrt{s}=3$\,TeV and $66.5\%$ at $10$\,TeV satisfy $p_T(H)<200$\,GeV for both Higgs bosons and fall in the resolved region. The boosted region, where both Higgs bosons exceed $200$\,GeV, accounts for $8.1\%$ of the signal at $3$\,TeV and $12.1\%$ at $10$\,TeV.

Signal and background samples are generated with \texttt{MadGraph5\_aMC@NLO}~\cite{Alwall:2014hca} using a UFO model with the modified trilinear coupling $\kappa_3$, with Higgs decays to $b\bar{b}$ handled by \texttt{MadSpin}~\cite{Artoisenet:2012st}. Parton showering is performed with \texttt{Pythia\,8}~\cite{Sjostrand:2014zea} and detector simulation with \texttt{Delphes\,3}~\cite{deFavereau:2013fsa} using the \texttt{delphes\_card\_MuonColliderDet.tcl} detector card.

The signal comprises the charged- and neutral-current processes,
\begin{align}
\mu^+\mu^- &\to HH\,\nu_\mu\bar\nu_\mu \to b\bar{b}b\bar{b}\,\nu_\mu\bar\nu_\mu, \label{eq:signal} \\
\mu^+\mu^- &\to HH\,\mu^+\mu^- \to b\bar{b}b\bar{b}\,\mu^+\mu^-,  \nonumber
\end{align}
introduced in Sec.~\ref{sec:hh_production}, where the muons in the second process are required to satisfy $|\eta_\mu|>2.5$ at generation and thus escape the detector acceptance.
Both processes therefore share the same experimental signature of four $b$-quarks with missing energy in the final state.
The charged current process dominates, with the neutral current channel contributing about $9\%$ ($11\%$) of the signal cross section at $\sqrt{s}=3$ ($10$)\,TeV at $\kappa_3=1$, and the two processes are treated as a single signal, denoted $HHX$, throughout the analysis.
Each signal and background sample contains $5\times10^5$ events at both $\sqrt{s}=3$ and $10$\,TeV.

To probe the $\kappa_3$ dependence of the signal, dedicated samples are generated at both $\sqrt{s}=3$ and $10$\,TeV with the trilinear coupling fixed to each value of the scan grid
\begin{equation}
\kappa_3 \in \{0.2,\, 0.4,\, 0.6,\, 0.8,\, 0.9,\, 1.0,\, 1.1,\, 1.2,\, 1.4,\, 1.6,\, 1.8\},
\label{eq:k3grid}
\end{equation}
with $10^5$ events per point. 
At $\sqrt{s}=10$\,TeV only, the grid is supplemented by additional points with a spacing of $\Delta\kappa_3=0.04$ over $\kappa_3\in[0.80,1.20]$,
\begin{equation}
\kappa_3 \in \{0.84,\ 0.88,\ 0.92,\ 0.96,\ 1.04,\ 1.08,\ 1.12,\ 1.16\},
\label{eq:k3grid_10tev}
\end{equation}
to finely sample the rapidly varying likelihood near $\kappa_3=1$.
The points of this finer scan, spanning $\kappa_3=0.80$ to $1.20$ in steps of $0.04$, are generated with $5\times10^5$ events each in the resolved region and $10^5$ events each in the boosted region, suppressing the template Monte Carlo statistical uncertainty in the region where the likelihood varies most rapidly.
The $\kappa_3$ discriminant is trained as a binary classifier between the $\kappa_3=0.4$ and $\kappa_3=1.6$ samples. To prevent overlap between the events used to train the classifiers and those used to build the likelihood templates and the Asimov dataset (Sec.~\ref{sec:statistics}), statistically independent samples ($10^5$ events each) are additionally generated at $\kappa_3\in\{0.4,\,1.0,\,1.6\}$.

In the resolved region, seven background processes are considered whose final states can mimic the signal topology of four jets and missing transverse energy,
\begin{align}
\text{Resolved region:}\quad & \mu^+\mu^- \to Hq\bar{q}\,\nu\bar\nu, \label{eq:bg_resolved}\\
& \mu^+\mu^- \to ZZ\,\nu\bar\nu, \nonumber\\
& \mu^+\mu^- \to W^+W^-\nu\bar\nu, \nonumber\\
& \mu^+\mu^- \to t\bar{t}\,\nu\bar\nu, \nonumber\\
& \mu^+\mu^- \to ZZ, \nonumber\\
& \mu^+\mu^- \to W^+W^-, \nonumber\\
& \mu^+\mu^- \to t\bar{t}. \nonumber
\end{align}
Here the Higgs boson decays to $b\bar{b}$ while the $Z$ and $W$ bosons and the top quark decay hadronically, and $\nu$ runs over all three neutrino flavors ($\nu_e,\nu_\mu,\nu_\tau$). The pair $q\bar q$ in $Hq\bar q\,\nu\bar\nu$ runs over all five flavors including $b$. Parton-level cuts of $p_T(q)>20$\,GeV, $|\eta_q|<2.5$, and $\Delta R(q,q)>0.4$ are applied to the final state quarks in the \texttt{MadGraph} generation, and for the di-boson samples the same cuts are propagated to the decay products of the bosons. No parton-level cuts are imposed on the top decay chain.
The four processes with $\nu\bar\nu$ in the final state share the same VBF topology as the signal and therefore resemble it kinematically more closely than the backgrounds without $\nu\bar\nu$.
Among them, $Hq\bar{q}\,\nu\bar\nu$ is the most signal-like, as it contains a genuine Higgs boson decaying to $b\bar b$.

In the boosted region, the signal samples are generated with a parton-level requirement of $p_T(H) > 200$\,GeV to prevent overlap with the resolved region. Eight background processes are considered in this topology:
\begin{align}
    \text{Boosted region:}\quad & \mu^+\mu^- \to HZ\,\nu\bar\nu, \label{eq:bg_boosted}\\
    & \mu^+\mu^- \to ZZ\,\nu\bar\nu, \nonumber\\
    & \mu^+\mu^- \to W^+W^-\,\nu\bar\nu, \nonumber\\
    & \mu^+\mu^- \to ZZ, \nonumber\\
    & \mu^+\mu^- \to W^+W^-, \nonumber\\
    & \mu^+\mu^- \to t\bar t, \nonumber\\
    & \mu^+\mu^- \to b\bar b, \nonumber\\
    & \mu^+\mu^- \to q\bar q. \nonumber
\end{align}
Here $q\bar q$ denotes a light quark pair ($q\neq b$). The $b\bar b$ final state is generated separately for sufficient sampling.
The $HZ\,\nu\bar\nu$ process takes the place of the resolved $Hq\bar{q}\,\nu\bar\nu$ background. Only its resonant component, on-shell $HZ$ production with $Z\to q\bar q$, can produce the collimated large-radius jets required in the boosted region, so the on-shell process is generated directly.
In the $HZ\,\nu\bar\nu$ sample the decays $H\to b\bar b$ and $Z\to q\bar q$ (including $b\bar b$) are forced at generation. For the remaining processes no decay is forced, and the bosons and top quarks decay inclusively in the parton shower, which allows the parton-level cuts to be placed directly on them. In the \texttt{MadGraph} generation, the $HZ\,\nu\bar\nu$ sample is required to have $p_T(H)>200$\,GeV and $p_T(Z)>150$\,GeV, the four di-boson samples to have $p_T(V)>200$\,GeV for each boson $V=W,Z$, and the two-body samples $t\bar t$, $b\bar b$, and $q\bar q$ to have $200\,\text{GeV}<p_T<850\,\text{GeV}$ on each final state top quark or quark. These cuts are employed to make the Monte Carlo samples focus on where the backgrounds can mimic the signal. The lower cuts correspond to the boosted region threshold. In the VBF signal much of the collision energy is carried away by the forward neutrinos, so the Higgs bosons populate the region just above $p_T=200$\,GeV, and the backgrounds are generated from the same threshold upward. 
On the other hand, in the $s$-channel two-body processes each final state particle instead carries about half of the collision energy over a broad angular distribution. Therefore, if an upper cut is not applied, most of the sample would sit far above the boosted selection window of Sec.~\ref{sec:presel_boosted} and be wasted, as will be demonstrated in the fat-jet $p_T$ distributions of Fig.~\ref{fig:preselection_boosted}. 
The upper cut at $850$\,GeV leaves a margin above that window for the parton-to-jet migration. The di-boson processes are dominated by $t$-channel exchange with steeply falling $p_T$ spectra, so no upper cut is needed there.

%--------------------------------------------------------------------------------
\section{Event Reconstruction and Selection}\label{sec:event_selection}

This section describes the jet reconstruction common to both kinematic regions and the pre-selection that defines the resolved and boosted samples.

Jets are reconstructed using the Valencia Linear Collider (VLC) algorithm~\cite{Boronat:2014hva,Boronat:2016tgd}, a sequential recombination algorithm designed for lepton collider environments.
The inter-particle and beam distances are defined as
\begin{equation}
    d_{ij} = \frac{2\min\!\left(E_i^{2\beta},\,E_j^{2\beta}\right)
              \left(1-\cos\theta_{ij}\right)}{R^{2}},
    \qquad
    d_{iB} = E_i^{2\beta}\sin^{2\gamma}\!\theta_{iB},
\label{eq:vlc_distance}
\end{equation}
where $E_i$ denotes the energy of the object $i$, $\theta_{ij}$ denotes the opening angle between the objects $i$ and $j$, $\theta_{iB}$ is the polar angle with respect to the beam axis, and $R$ is the radius parameter. We adopt $\beta = \gamma = 1$ throughout, following the default VLC configuration of the muon collider Delphes card.
We consider exclusive clustering, forcing a fixed number of jets per event. Each of the two kinematic regions defined in Sec.~\ref{sec:signal_background} uses its own jet clustering. The resolved region is clustered into exactly four jets with $R=0.5$ (the Delphes branch \texttt{VLCjetR05N4}), and the boosted region into exactly two large-radius jets with $R=1.0$ (\texttt{VLCjetR10N2}).
Flavor tagging is applied to the jets of both regions at a common working point of the \texttt{MuonColliderDet\_BTag\_70.tcl} card, corresponding to a per-jet $b$-tagging efficiency of $70\%$ (taken flat in $p_T$ and $\eta$), a $c$-jet mis-tag rate ranging from $\simeq1\%$ in the barrel to $\simeq10\%$ in the forward region ($|\eta|\gtrsim2.1$), and a light flavor mis-tag rate at the $0.1$--$2\%$ level.

\subsection{Resolved region}\label{sec:presel_resolved}

Events in the resolved region are selected by the following pre-selection criteria:
\begin{itemize}
    \item four jets with $R=0.5$, $p_T(j)>20$\,GeV, and $|\eta_j|<2.5$;
    \item no isolated electron or muon with $p_T>10$\,GeV and $|\eta|<2.5$ (lepton veto);
    \item both Higgs candidates satisfying $p_T(H_{1,2})<200$\,GeV.
\end{itemize}

For the pre-selection, the Higgs candidates are reconstructed by minimizing, over the three possible two-jet pairings, the mass window compatibility variable~\cite{ATLAS:2022hwc}
\begin{equation}
X_{HH}=\sqrt{\left(\frac{m_{H_1}-M_1}{0.1\,m_{H_1}}\right)^{2}+\left(\frac{m_{H_2}-M_2}{0.1\,m_{H_2}}\right)^{2}},
\label{eq:XHH}
\end{equation}
where $M_1$ and $M_2$ are reference masses and the widths are taken to be $10\%$ of the candidate masses, both following Ref.~\cite{ATLAS:2022hwc}. We use the values $M_1=120$\,GeV and $M_2=110$\,GeV in the resolved channel, while $M_1=124$\,GeV and $M_2=115$\,GeV in the boosted channel, where the same variable is evaluated with the large-radius jet masses in Sec.~\ref{sec:ml_boosted}. 
They lie below $m_H$ because detector effects, the energy carried away by neutrinos in the $b$-hadron decays, and out-of-cone radiation all shift the reconstructed candidate masses downward. 
The final jet-to-Higgs assignment used in the analysis is instead performed by the SPANet network of Sec.~\ref{sec:ml_resolved}.

\begin{table}[t]
    \centering
    \setlength{\tabcolsep}{5pt}
    % \footnotesize
    \begin{tabular}{lccccccc}
    \hline\hline
    & \multicolumn{3}{c}{$\sqrt{s}=3$\,TeV} & & \multicolumn{3}{c}{$\sqrt{s}=10$\,TeV}\\
    \cline{2-4}\cline{6-8}
    Process & $\sigma_{\rm gen}$ [fb] & $\epsilon_{\rm pre}$ [\%] & $N_{\rm pre}$
        & & $\sigma_{\rm gen}$ [fb] & $\epsilon_{\rm pre}$ [\%] & $N_{\rm pre}$ \\
    \hline
    $HHX$                             & $0.312$ & $42.4$ & $132$    & & $1.273$  & $27.2$ & $3{,}467$   \\
    \hline
    $Hq\bar{q}\nu\bar\nu$             & $3.164$ & $45.2$ & $1{,}431$  & & $8.520$  & $32.9$ & $27{,}989$  \\
    $ZZ\nu\bar\nu$                    & $12.16$ & $45.3$ & $5{,}514$  & & $23.93$  & $36.7$ & $87{,}860$  \\
    $W^+W^-\nu\bar\nu$                & $19.52$ & $47.6$ & $9{,}292$  & & $33.97$  & $40.6$ & $137{,}855$ \\
    $t\bar{t}\nu\bar\nu$              & $2.228$ & $58.8$ & $1{,}310$  & & $6.982$  & $45.4$ & $31{,}726$  \\
    $ZZ$                              & $0.755$ & $14.4$ & $109$    & & $0.0004$ & $3.7$  & $0.1$ \\
    $W^+W^-$                          & $9.191$ & $22.2$ & $2{,}043$ & & $0.0026$ & $4.4$  & $1.1$ \\
    $t\bar{t}$                        & $7.792$ & $1.6$  & $122$    & & $0.704$  & $0.3$  & $23.4$   \\
    \hline
    Total background                  & $54.81$ & ---    & $19{,}821$ & & $74.11$  & ---    & $285{,}455$ \\
    \hline\hline
    \end{tabular}
    \caption{Production cross section $\sigma_{\rm gen}$, pre-selection efficiency $\epsilon_{\rm pre}$, and expected yield $N_{\rm pre}$ in the resolved region at $\sqrt{s}=3$\,TeV ($\mathcal{L}=1$\,ab$^{-1}$) and $\sqrt{s}=10$\,TeV ($\mathcal{L}=10$\,ab$^{-1}$). Here $\sigma_{\rm gen}$ is evaluated with the forced decay modes and the parton-level quark cuts of Sec.~\ref{sec:signal_background}, so it already includes the hadronic branching fractions. The pre-selection criteria are defined in the text. The signal $HHX$ denotes $\mu^+\mu^-\to HH+X$ with $X\in\{\nu_\mu\bar\nu_\mu,\,\mu^+\mu^-\}$, evaluated at $\kappa_3=1$ (SM).}
    \label{tab:preselection_resolved}
\end{table}

\begin{table}[t]
    \centering
    \setlength{\tabcolsep}{5pt}
    % \footnotesize
    \begin{tabular}{cccccccc}
    \hline\hline
    & \multicolumn{3}{c}{$\sqrt{s}=3$\,TeV} & & \multicolumn{3}{c}{$\sqrt{s}=10$\,TeV}\\
    \cline{2-4}\cline{6-8}
    $\kappa_3$ & $\sigma_{\rm gen}$ [fb] & $\epsilon_{\rm pre}$ [\%] & $N_{\rm pre}$
           & & $\sigma_{\rm gen}$ [fb] & $\epsilon_{\rm pre}$ [\%] & $N_{\rm pre}$ \\
    \hline
    $0.2$ & $0.567$ & $45.5$ & $258$  & & $1.886$ & $30.8$ & $5{,}811$ \\
    $0.4$ & $0.485$ & $44.9$ & $218$  & & $1.689$ & $30.1$ & $5{,}086$ \\
    $0.6$ & $0.417$ & $43.8$ & $182$  & & $1.527$ & $29.3$ & $4{,}479$ \\
    $0.8$ & $0.358$ & $43.1$ & $154$  & & $1.383$ & $28.3$ & $3{,}913$ \\
    $0.9$ & $0.333$ & $42.8$ & $143$  & & $1.324$ & $27.9$ & $3{,}697$ \\
    $1.0$ & $0.312$ & $42.4$ & $132$  & & $1.273$ & $27.2$ & $3{,}467$ \\
    $1.1$ & $0.294$ & $41.9$ & $123$  & & $1.225$ & $27.0$ & $3{,}308$ \\
    $1.2$ & $0.278$ & $41.3$ & $115$  & & $1.184$ & $26.5$ & $3{,}135$ \\
    $1.4$ & $0.255$ & $40.6$ & $104$  & & $1.127$ & $26.0$ & $2{,}925$ \\
    $1.6$ & $0.245$ & $40.2$ & $98$ & & $1.092$ & $25.5$ & $2{,}783$ \\
    $1.8$ & $0.246$ & $40.0$ & $98$ & & $1.085$ & $25.6$ & $2{,}774$ \\
    \hline\hline
    \end{tabular}
    \caption{Signal cross section $\sigma_{\rm gen}$ for $\mu^+\mu^-\to HH+X\to b\bar{b}b\bar{b}+X$ in the resolved region, pre-selection efficiency $\epsilon_{\rm pre}$, and expected yield $N_{\rm pre}$ as a function of $\kappa_3$, at $\sqrt{s}=3$\,TeV ($\mathcal{L}=1$\,ab$^{-1}$) and $\sqrt{s}=10$\,TeV ($\mathcal{L}=10$\,ab$^{-1}$). The pre-selection criteria are defined in the text.}
    \label{tab:preselection_signal_resolved}
\end{table}

Table~\ref{tab:preselection_resolved} lists the production cross section $\sigma_{\rm gen}$, the pre-selection efficiency $\epsilon_{\rm pre}$, and the expected yield $N_{\rm pre}$ for each signal and background process at $\sqrt{s}=3$\,TeV ($\mathcal{L}=1$\,ab$^{-1}$) and $\sqrt{s}=10$\,TeV
($\mathcal{L}=10$\,ab$^{-1}$).
Here and in the following tables, $\sigma_{\rm gen}$ denotes the cross section at the generation level, including the decays forced in \texttt{MadGraph} and the parton-level cuts of Sec.~\ref{sec:signal_background}.
The pre-selection retains the signal and the $\nu\bar\nu$-accompanied backgrounds with comparable efficiencies, while the two-body backgrounds are suppressed by a further factor of two to three at $3$\,TeV and by an order of magnitude at $10$\,TeV, where their cross sections have themselves collapsed. The surviving background, dominated by the VBF-type processes, still exceeds the signal by two orders of magnitude, which motivates the machine learning analysis of Sec.~\ref{sec:ml_likelihood}.
Table~\ref{tab:preselection_signal_resolved} gives the
analogous quantities for the signal as a function of $\kappa_3$.
Across the scanned grid the signal yield varies by a factor of $2.6$ at $3$\,TeV and $2.1$ at $10$\,TeV between $\kappa_3=0.2$ and $1.8$, reflecting the destructive interference of Sec.~\ref{sec:hh_production}, and $\epsilon_{\rm pre}$ decreases mildly with $\kappa_3$ as the suppressed threshold contribution hardens the $p_T(H)$ spectrum.

\subsection{Boosted region}\label{sec:presel_boosted}

\begin{figure}[t]
    \centering
    \includegraphics[width=\linewidth]{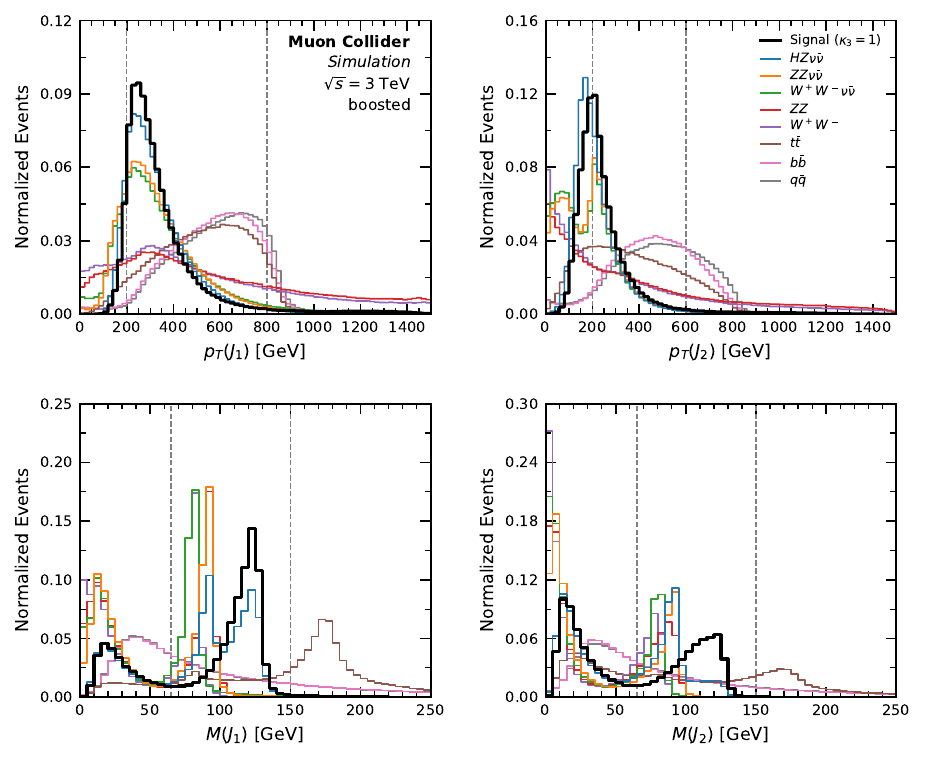}
    \caption{Transverse momentum and mass distributions of the leading and subleading large-radius jets in the boosted region at $\sqrt{s}=3$\,TeV, after the parton-level generation cuts of Sec.~\ref{sec:signal_background} and the requirement of two fat jets, and before the remaining pre-selection cuts (vertical dashed lines). The black line shows the signal ($\kappa_3=1$) and each background process is shown separately, with every distribution normalized to unity.}
    \label{fig:preselection_boosted}
\end{figure}

In the boosted region, the decay products of each Higgs boson are collimated and merge into a single large-radius jet, denoted by 
$J$ in the following. The signature of this region is therefore a pair of Higgs candidate jets clustered with $R=1.0$, each carrying a large transverse momentum, $p_T(J)>200$\,GeV, and a jet mass close to the Higgs boson mass.

Fig.~\ref{fig:preselection_boosted} presents the transverse momenta and masses of the leading and subleading fat jets, separately for the signal and for each background process. The following pre-selection cuts are applied:
\begin{itemize}
    \item two fat jets with $R=1.0$ and $|\eta(J)|<2.5$;
    \item $200~\text{GeV}< p_T(J_{1})<800~\text{GeV}$;
    \item $200~\text{GeV}< p_T(J_{2})<600~\text{GeV}$;
    \item $65~\text{GeV}< M(J_{1,2})<150~\text{GeV}$;
\end{itemize}
where $J_1$ and $J_2$ represent the leading jet and the subleading jet
arranged by the $p_T$, respectively.
Unlike in the resolved region, no veto is applied on isolated leptons. In the boosted topology most leptons arise from heavy flavor decays inside the fat jets and are retained as low-level classifier inputs (Sec.~\ref{sec:ml_boosted}).

The lower $p_T$ cuts separate the boosted from the resolved region, while the upper $p_T$ cuts remove backgrounds, in particular the two-body final states, which accumulate at high transverse momentum. The upper $M(J)$ cut suppresses jets from boosted top quarks, as clearly indicated by the $t\bar t$ background $M(J_1)$ and $M(J_2)$, which peak near the top-quark mass.
The small bump of signal events in the low mass region originates from mis-clustered Higgs decay products, which would degrade the subsequent ML classification. The lower $M(J)$ cut removes these events and also parts of the background. The $HZ\,\nu\bar\nu$ background shows similar resonant structure to the signal. The $M(J_1)$ distribution peaks near both $m_H$ and $m_Z$, depending on which boson jet leads in transverse momentum, while $M(J_2)$ peaks near $m_Z$. Both peaks lie inside the mass window, so this background largely survives the pre-selection.

Table~\ref{tab:preselection_boosted} lists $\sigma_{\rm gen}$, $\epsilon_{\rm pre}$, and $N_{\rm pre}$ for the boosted region. The largest background contribution comes from di-boson production, owing to its large cross section, but its kinematic distributions differ markedly from the signal, as shown in Sec.~\ref{sec:ml_boosted}, allowing the subsequent machine learning classifiers to suppress it. The $HZ\,\nu\bar\nu$ process, whose resonant structure survives the mass window as noted above, is instead expected to be the most difficult background to reject.
After the pre-selection the background exceeds the signal by more than three orders of magnitude at $3$\,TeV, where only about six signal events are expected, and this scarcity will dictate the simplified statistical treatment of Sec.~\ref{sec:statistics}. As in the resolved region, the two-body backgrounds collapse at $10$\,TeV and the $\nu\bar\nu$-accompanied processes dominate.
The two signal processes, generated separately in the boosted region, have nearly identical pre-selection efficiencies ($29.7\%$ and $28.9\%$ at $3$\,TeV, $22.7\%$ and $24.4\%$ at $10$\,TeV), and are therefore merged into the single $HHX$ entry in Table~\ref{tab:preselection_boosted}, with the neutral-current share remaining at the $8$--$13\%$ level after the selection.

\begin{table}[t]
    \centering
    \setlength{\tabcolsep}{5pt}
    % \footnotesize
    \begin{tabular}{lccccccc}
    \hline\hline
    & \multicolumn{3}{c}{$\sqrt{s}=3$\,TeV} & & \multicolumn{3}{c}{$\sqrt{s}=10$\,TeV}\\
    \cline{2-4}\cline{6-8}
    Process & $\sigma_{\rm gen}$ [fb] & $\epsilon_{\rm pre}$ [\%] & $N_{\rm pre}$ & & $\sigma_{\rm gen}$ [fb] & $\epsilon_{\rm pre}$ [\%] & $N_{\rm pre}$ \\
    \hline
    $HHX$                          & $0.0208$ & $29.6$ & $6.15$ & & $0.128$ & $23.0$ & $293.42$ \\
    \hline
    $HZ\nu\bar\nu$                              & $0.5196$   & $26.6$  & $138.22$   & & $2.8644$   & $13.8$   & $3{,}953$ \\
    $ZZ\nu\bar\nu$                              & $6.469$   & $24.2$  & $1{,}567$  & & $39.84$   & $14.8$   & $59{,}022$ \\
    $W^+W^-\nu\bar\nu$                          & $12.23$   & $20.0$  & $2{,}451$  & & $70.54$   & $11.6$   & $81{,}832$ \\
    $ZZ$                                        & $9.617$   & $12.2$  & $1{,}168$  & & $1.336$   & $0.0036$ & $0.48$  \\
    $W^+W^-$                                    & $150.7$   & $10.1$  & $15{,}287$ & & $22.01$   & $0.0014$ & $3.08$  \\
    $t\bar t$                                   & $4.424$   & $7.2$   & $316.25$   & & $0.0354$  & $3.7$    & $12.97$  \\
    $b\bar b$                                   & $2.326$   & $7.8$   & $181.78$   & & $0.0187$  & $2.1$    & $3.87$   \\
    $q\bar q$ ($q\neq b$)                       & $13.45$   & $7.1$   & $949.81$   & & $0.108$   & $1.9$    & $20.50$  \\
    \hline
    Total background                            & $199.7$   & ---     & $22{,}059$ & & $136.8$  & ---      & $144{,}848$ \\
    \hline\hline
    \end{tabular}
    \caption{Production cross section $\sigma_{\rm gen}$, pre-selection efficiency $\epsilon_{\rm pre}$, and expected yield $N_{\rm pre}$ in the boosted region at $\sqrt{s}=3$\,TeV ($\mathcal{L}=1$\,ab$^{-1}$) and $\sqrt{s}=10$\,TeV ($\mathcal{L}=10$\,ab$^{-1}$). Here the parton-level cuts of Sec.~\ref{sec:signal_background} are applied directly to the bosons and top quarks, which then decay inclusively in the shower, so $\sigma_{\rm gen}$ includes the hadronic branching fractions only for the $HZ\nu\bar\nu$ sample and for the other processes they are absorbed into $\epsilon_{\rm pre}$. The pre-selection criteria are defined in the text. The signal $HHX$ denotes $\mu^+\mu^-\to HH+X$ with $X\in\{\nu_\mu\bar\nu_\mu,\,\mu^+\mu^-\}$, evaluated at $\kappa_3=1$ (SM).}
    \label{tab:preselection_boosted}
\end{table}

\begin{table}[t]
    \centering
    \setlength{\tabcolsep}{5pt}
    % \footnotesize
    \begin{tabular}{lccccccc}
    \hline\hline
               & \multicolumn{3}{c}{$\sqrt{s}=3$\,TeV}              & & \multicolumn{3}{c}{$\sqrt{s}=10$\,TeV}   \\
    \cline{2-4}\cline{6-8}
    $\kappa_3$ & $\sigma_{\rm gen}$ [fb] & $\epsilon_{\rm pre}$ [\%] & $N_{\rm pre}$ & & $\sigma_{\rm gen}$ [fb] & $\epsilon_{\rm pre}$ [\%] & $N_{\rm pre}$ \\
    \hline
    0.2        & 0.0455       & 31.4                    & 14.29     & & 0.221        & 27.0                    & 596.25    \\
    0.4        & 0.0375       & 31.0                    & 11.62     & & 0.192        & 26.2                    & 501.76    \\
    0.6        & 0.0310       & 30.6                    & 9.48      & & 0.166        & 25.2                    & 418.39    \\
    0.8        & 0.0251       & 30.4                    & 7.64      & & 0.145        & 24.5                    & 353.79    \\
    0.9        & 0.0229       & 29.9                    & 6.85      & & 0.136        & 23.8                    & 323.26    \\
    1.0        & 0.0208       & 29.6                    & 6.15      & & 0.128        & 23.0                    & 293.42    \\
    1.1        & 0.0190       & 29.0                    & 5.52      & & 0.121        & 22.3                    & 269.32    \\
    1.2        & 0.0176       & 28.6                    & 5.01      & & 0.115        & 21.8                    & 249.96    \\
    1.4        & 0.0154       & 27.4                    & 4.22      & & 0.106        & 20.6                    & 218.19    \\
    1.6        & 0.0145       & 26.5                    & 3.84      & & 0.101        & 19.6                    & 197.46    \\
    1.8        & 0.0147       & 25.7                    & 3.78      & & 0.100        & 19.2                    & 192.64    \\
    \hline\hline
    \end{tabular}
    \caption{Signal cross section $\sigma_{\rm gen}$ for $\mu^+\mu^-\to HH+X\to b\bar{b}b\bar{b}+X$, generated with the $p_T(H)>200$\,GeV requirement of Sec.~\ref{sec:signal_background}, pre-selection efficiency $\epsilon_{\rm pre}$, and expected yield $N_{\rm pre}$ as a function of $\kappa_3$, at $\sqrt{s}=3$\,TeV ($\mathcal{L}=1$\,ab$^{-1}$) and $\sqrt{s}=10$\,TeV ($\mathcal{L}=10$\,ab$^{-1}$). The cross section and $N_{\rm pre}$ are summed over the charged current and neutral current signal channels.}
    \label{tab:preselection_signal_boosted}
\end{table}

Table~\ref{tab:preselection_signal_boosted} gives the analogous quantities as a function of $\kappa_3$. The ratio of the $10$\,TeV to the $3$\,TeV signal cross section grows from $4.9$ at $\kappa_3=0.2$ to $7.0$ at $\kappa_3=1.6$, whereas the corresponding resolved ratio in Table~\ref{tab:preselection_signal_resolved} only grows from $3.3$ to $4.5$. This confirms the expectation that the boosted region becomes increasingly important as the center-of-mass energy grows.

%--------------------------------------------------------------------------------
\section{Machine Learning Models and Likelihood Analysis}\label{sec:ml_likelihood}

This section introduces the topological descriptors used as input features, describes the machine learning models of the two regions, and defines the likelihood analysis from which the results of Sec.~\ref{sec:results} are obtained.

\subsection{Topological data analysis}\label{sec:tda}

Topological data analysis (TDA)~\cite{Edelsbrunner:2002} is a recently developed technique for characterizing the topological structure, i.e.\ the ``shape'', of a data set.
In collider physics, persistent homology has been applied to the internal topology of jets~\cite{Li:2020jdb,Mittal:2026tss} and to event-level observables~\cite{Beuria:2023seb}, and has recently been combined with machine learning in a muon collider study~\cite{Beuria:2025avt}.
We use five TDA descriptors $\{H_0, H_1, S_0, S_1, LB_1\}$ to summarize the event-level $(\eta,\phi)$ topology of the detector energy flow through persistent homology.

These five variables provide a global topological description of each event and enter the machine learning models as high-level input features. We build a Vietoris--Rips filtration on the cylindrical inter-particle angular distance\footnote{The filtration is built from all detector-level EFlow constituents of the event with $E_T>0$.}
\begin{equation}
\rho_{ij}\;=\;\sqrt{(\Delta\eta_{ij})^2 \;+\; \bigl(2\sin(\Delta\phi_{ij}/2)\bigr)^2}\,,
\label{eq:tda_distance}
\end{equation}
and extract the birth/death pairs $(b_k,d_k)$ of the zero- and one-dimensional homology classes ($H_0$ for connected components and $H_1$ for one-dimensional loops) with the \texttt{Ripser}~\cite{Bauer_2021} library. Writing $\ell_k\equiv d_k-b_k$ as the persistence ``lifetime'' of class $k$, the descriptors are defined as
\begin{equation}
\begin{aligned}
H_0 &= \!\!\sum_{k\in{\rm dgm}_0,\,d_k<\infty}\!\!\!\ell_k, &
H_1 &= \sum_{k\in{\rm dgm}_1}\!\!\ell_k, &
LB_1 &= \sum_{k\in{\rm dgm}_1}\!\!b_k\,\ell_k,\\
S_0 &= -\!\!\sum_{k\in{\rm dgm}_0,\,d_k<\infty}\!\!\!\frac{\ell_k}{H_0}\log_2\!\frac{\ell_k}{H_0}, &
S_1 &= -\!\!\sum_{k\in{\rm dgm}_1}\!\!\frac{\ell_k}{H_1}\log_2\!\frac{\ell_k}{H_1}, & &
\end{aligned}
\label{eq:tda_descriptors}
\end{equation}
where ${\rm dgm}_0$ and ${\rm dgm}_1$ denote the persistence diagrams, the multisets of birth/death pairs of the $H_0$ and $H_1$ classes, and the infinite-lifetime $H_0$ class (the single connected component surviving to the end of the filtration) is excluded from the sums. The zero-dimensional analog $LB_0$ is omitted because every $H_0$ class is born at $b_k=0$ and it would therefore carry no information. Descriptors of an empty diagram, e.g.\ $H_1$, $S_1$, and $LB_1$ for events with no one-dimensional loops, are set to zero.

Physically, $H_0$ measures the total angular scale over which separate energy clusters merge into a single one. It grows when the energy deposition is angularly spread out (broad, isotropic events) and is small for compact, low-multiplicity ones. $H_1$ counts and weights the angular ``voids'' enclosed by rings of particles, and is enhanced in topologies with several well-separated energy clusters, making it sensitive to back-to-back, multi-pronged signatures such as a $4b$ final state from a $HH$ system. $S_0$ and $S_1$ are Shannon entropies of the corresponding normalized lifetime spectra. They are large when many homology features of comparable persistence coexist (multi-scale geometry) and small when a single feature dominates. $LB_1$ weights each one-dimensional loop by the angular scale at which it is born, emphasizing the large, late-forming voids characteristic of sparse, isolated clusters. The five descriptors together provide a frame-invariant geometric summary of the $(\eta,\phi)$ energy flow that is invariant under longitudinal boosts and azimuthal rotations and complementary to the kinematic high-level variables.

\subsection{Resolved region}\label{sec:ml_resolved}

As outlined in Sec.~\ref{sec:introduction}, two networks are trained in each region. The first, $\mathcal{D}_{\rm HH}$, is trained to separate the SM signal from the total background. The second, $\mathcal{D}_{\kappa_3}$, is trained on the binary task of discriminating $\kappa_3=0.4$ from $\kappa_3=1.6$, acting as a $\kappa_3$ shape discriminant. The two values are chosen symmetric about the SM point and far enough apart that a clear shape difference develops in the $\kappa_3$-sensitive observables, most visibly in the $m_{HH}$ spectrum. The trained classifier is applied to all $\kappa_3$ hypotheses in the likelihood scan. In the resolved region the four jets must first be assigned to the two Higgs candidates, and a dedicated SPANet model performs this pairing.

\begin{figure}[ht]
    \centering
    \includegraphics[width=\linewidth]{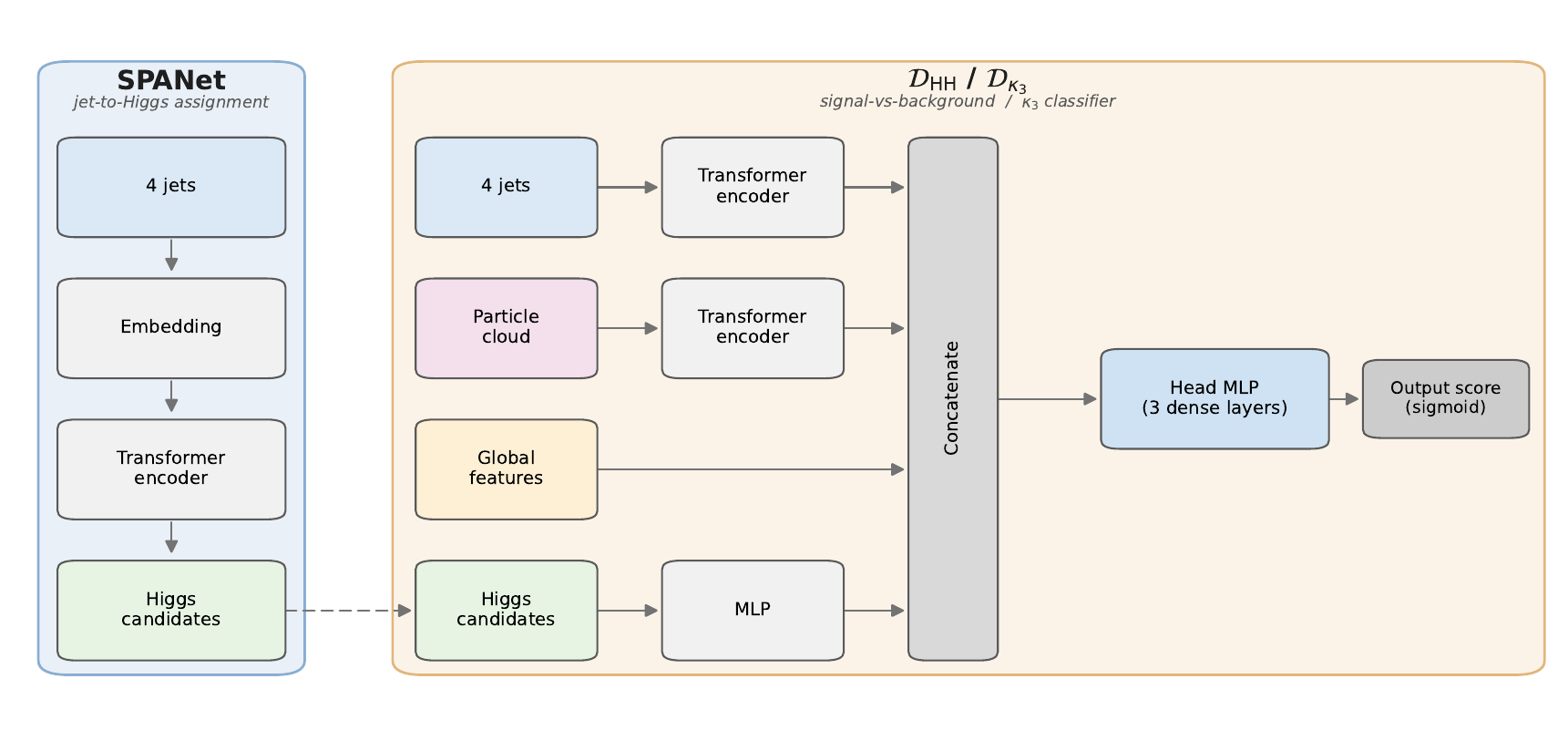}
    \caption{Architecture of the SPANet jet-to-Higgs assignment network (left) and the shared $\mathcal{D}_{\rm HH}$/$\mathcal{D}_{\kappa_3}$ multi-stream classifier (right) used in the resolved analysis. SPANet embeds the four jets and applies a self-attention (transformer) encoder to assign them to two Higgs candidates. These candidates (dashed arrow) form the fourth $\mathcal{D}_{\rm HH}$/$\mathcal{D}_{\kappa_3}$ input stream. The other three are the four jets, the low-level particle cloud, and the global (high-level $+$ TDA) features. Each stream is processed by its own block, a multi-layer perceptron or a transformer encoder, and the outputs are concatenated and passed through a three-layer head to a sigmoid output score. $\mathcal{D}_{\rm HH}$ and $\mathcal{D}_{\kappa_3}$ share this architecture but are trained separately to separate the signal from the background and $\kappa_3{=}0.4$ from $\kappa_3{=}1.6$.}
    \label{fig:network_resolved}
\end{figure}

The assignment classifies each event into one of the three possible jet pairings, the three ways of splitting four jets into two pairs. We perform it with an attention-based network~\cite{Vaswani:2017ovl} inspired by the SPANet (Symmetry Preserving Attention Network) approach~\cite{Shmakov:2021qdz} (left block of Fig.~\ref{fig:network_resolved}), trained in a supervised manner on truth-matched signal events rather than relying on a fixed kinematic rule. The original SPANet addresses the general jet--parton assignment problem with symmetric tensor attention heads. Our network instead classifies the three possible pairings of the four jets directly and keeps only the symmetry-preserving structure. Each jet is represented by six features,
\begin{equation*}
\bigl[\,\ln p_T,\ \ \eta,\ \ \sin\phi,\ \ \cos\phi,\ \ \ln(1+m/m_0),\ \ b\text{-tag}\,\bigr],
\end{equation*}
where the reference mass $m_0=5$\,GeV keeps the mass input dimensionless, smooth at $m\to 0$, and $\mathcal{O}(1)$.
The four jets are processed by a multi-head self-attention encoder which carries no positional encoding and is invariant to their input ordering. Each Higgs candidate is represented by the mean of its two jet embeddings, and the score of each di-jet partition is built from $H_1\leftrightarrow H_2$-symmetric combinations of the two candidate embeddings (their sum and the magnitude of their difference), so that the assignment is invariant under the intra-candidate jet swap, the $H_1\leftrightarrow H_2$ exchange, and the ordering of the input jets.

\begin{figure}[ht]
    \centering
    \includegraphics[width=\textwidth]{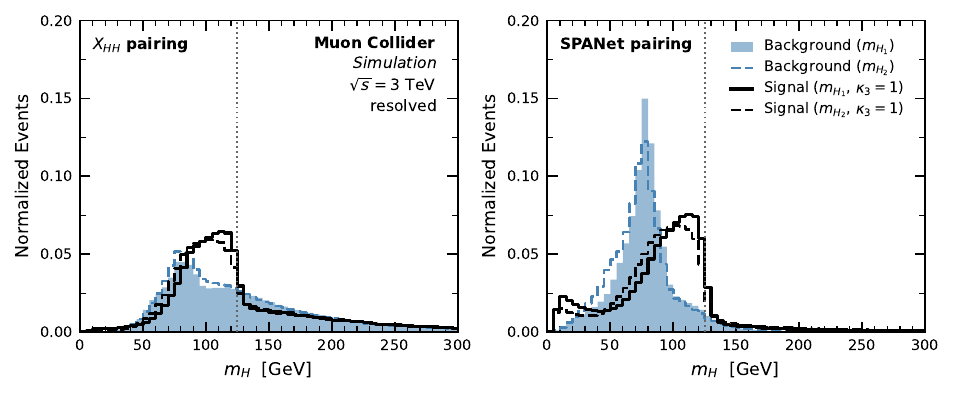}
    \caption{Higgs candidate mass distributions $m_{H_1}$ (solid) and $m_{H_2}$ (dashed) under the traditional $X_{HH}$ pairing (left) and the SPANet pairing (right) at $\sqrt{s}=3$\,TeV. The black line shows the SM signal ($\kappa_3=1$) and the blue filled histograms show the inclusive background. The vertical dotted gray line marks $m_H=125$\,GeV.}
    \label{fig:pairing_comparison}
\end{figure}

To illustrate the impact of the jet pairing choice quantitatively, we compare in Fig.~\ref{fig:pairing_comparison} the two Higgs candidate masses, $m_{H_1}$ (leading $p_T$, solid) and $m_{H_2}$ (subleading $p_T$, dashed), obtained under the traditional $X_{HH}$ pairing of Eq.~(\ref{eq:XHH}) and under the SPANet pairing used in our headline analysis, at $\sqrt{s}=3$\,TeV.
By construction the $X_{HH}$ pairing biases every event towards the Higgs mass window, so the signal and inclusive background distributions come out broad and strongly overlapping (left panel), with the genuine signal peak diluted by mis-assigned combinations.
The SPANet pairing instead produces a markedly cleaner separation (right panel). Its assignment accuracy, the fraction of held-out truth-matched events in which both Higgs candidates are correctly assembled, is $96\%$ at $3$\,TeV and $94\%$ at $10$\,TeV. Correct jet assignment enables the signal to peak relatively closer to $m_H\simeq 125$\,GeV with fewer events leaking to $m_H>125$\,GeV.
The backgrounds, no longer forced into the Higgs mass window, peak more sharply at their physical origins near $m_W\simeq 80$\,GeV and $m_Z\simeq 91$\,GeV.

Both SPANet pairings leave a small secondary shoulder near $m_{H_1}\!\lesssim\! 30$\,GeV in the signal distribution. We verified that this feature is not a pairing failure but a reconstruction-stage artifact. Most shoulder events have at least one parton-level $b$-quark falling outside the jet kinematic window ($p_T<20$\,GeV or $|\eta|>2.5$), and in most of the remaining cases at least one $b$-quark lies at $\Delta R>0.5$ from every reconstructed jet. The shoulder is therefore a $b$-quark loss at the four-jet reconstruction step rather than a deficiency of the pairing algorithm.

The SPANet pairing is followed by the two event-level classifiers $\mathcal{D}_{\rm HH}$ and $\mathcal{D}_{\kappa_3}$, which share an identical multi-stream architecture and differ only in their training target. As the right block of Fig.~\ref{fig:network_resolved} shows, four input streams are processed in parallel:
\begin{itemize}
\item \textbf{Jets}: for each of the four leading jets, $\ln p_T^{\,j}$, $\eta_j$, $\sin\phi_j$, $\cos\phi_j$, $\ln(1+m_j/m_0)$, $|\Delta\phi(j,E_T^{\rm miss})|$, and the jet $b$-tag (separate embedding).
\item \textbf{Higgs candidates}: for each of the two Higgs candidates from the SPANet assignment, $\ln p_T^{\,H}$, $\eta_H$, $\sin\phi_H$, $\cos\phi_H$, $\ln m_H$, the number of $b$-tagged jets $N_b^{H}$, and $\Delta R_{jj}$.
\item \textbf{Global features}: $m_{HH}$, $p_T^{HH}$, $X_{HH}$ of Eq.~(\ref{eq:XHH}), $\Delta\phi(HH,E_T^{\rm miss})$, $\Delta R_{H_1H_2}$, the total $b$-tag multiplicity, $E_T^{\rm miss}$, the scalar sum of energies, and the five TDA descriptors $H_0$, $H_1$, $S_0$, $S_1$, $LB_1$ (Sec.~\ref{sec:tda}).
\item \textbf{Particle cloud}: for the ten leading EFlow~\cite{deFavereau:2013fsa} constituents of each jet, $p_T^{\,i}/p_T^{\mathrm{jet}}$, $\Delta\eta_i$, and $\Delta\phi_i$ relative to the parent jet axis, and a particle type code (embedding), with electron, muon, photon, neutral hadron, and charged hadron labeled $1$--$5$.
\end{itemize}
The four stream representations are concatenated and passed through a head multi-layer perceptron (MLP) of three Dense$+$Dropout layers, followed by a Dense$(1)+$sigmoid output. A similar design, in which attention-based encoders process different levels of event information separately before being merged, has been applied to $HH\to b\bar bb\bar b$ searches at the LHC~\cite{Hammad:2023sbd}.

The SPANet model is implemented in \texttt{PyTorch}~\cite{NEURIPS2019_bdbca288} and the two classifiers in \texttt{Keras}~\cite{chollet2015keras}.
The depth and width of each stream block and of the head, together with the training hyper-parameters, are optimized with \texttt{Optuna}~\cite{akiba2019optuna}, maximizing the validation-set AUC. The two resolved-region classifiers are trained on a $70\%/15\%/15\%$ training/validation/test split of the simulated samples, and all performance figures and likelihood templates below are built from the held-out test partition only, scaled to the target luminosity. The optimized configuration and the full training code are available in a public repository.\footnote{\url{https://github.com/LSJ957/muc-hh-kappa3}}

\begin{figure}[t]
    \centering
    \includegraphics[width=\textwidth]{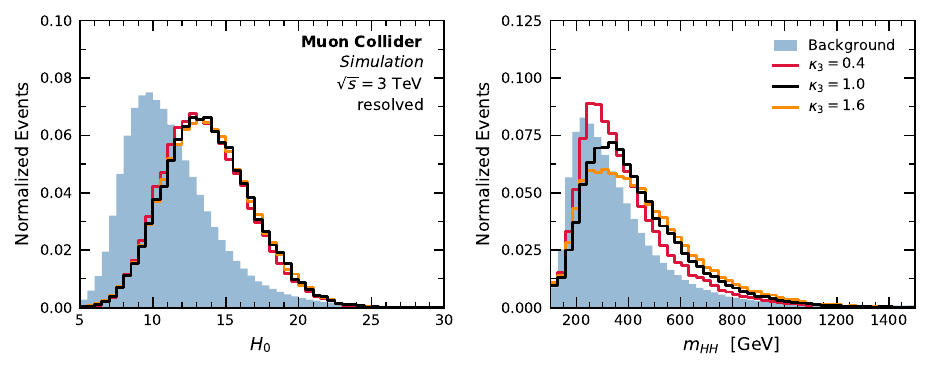}
    \caption{Representative resolved region observables at $\sqrt{s}=3$\,TeV for three signal hypotheses, $\kappa_3=0.4$ (red), $\kappa_3=1.0$ (black), and $\kappa_3=1.6$ (orange), and the inclusive background (filled blue). The left panel shows the TDA variable $H_0$, which separates signal from background but is nearly independent of $\kappa_3$. The right panel shows the four-jet invariant mass $m_{HH}$, which varies with $\kappa_3$ and discriminates between trilinear coupling hypotheses.}
    \label{fig:kinematics_resolved}
\end{figure}

Fig.~\ref{fig:kinematics_resolved} shows two representative resolved region observables at $\sqrt{s}=3$\,TeV for three signal hypotheses, $\kappa_3=0.4$ (red), $1.0$ (black), and $1.6$ (orange), and the inclusive background (filled blue). The left panel shows the persistent homology variable $H_0$, which separates the $HH\to b\bar{b}b\bar{b}$ signal from the bulk of the background because the four $b$-jets of the signal form a richer multi-prong topology. It is essentially independent of $\kappa_3$, since every hypothesis shares the same $HH\to b\bar{b}b\bar{b}$ final state and hence the same event topology. The right panel shows the four-jet invariant mass $m_{HH}$, which instead varies with $\kappa_3$ and discriminates between trilinear coupling hypotheses. This dependence follows from the trilinear amplitude $WW/ZZ\to H^*\to HH$, whose propagator enhances the threshold region $m_{HH}\gtrsim 2m_H$.
The cross section minimum near $\kappa_3\simeq 1.7$ arises from the destructive interference of this amplitude with the non-trilinear diagrams of Fig.~\ref{fig:feynman_diagrams}, i.e.\ the quartic $HHVV$ and double-$HVV$ contributions. The interference is strongest at threshold. Among the three $\kappa_3$ values, $\kappa_3=1.6$ lies closest to this minimum, so its threshold region is most suppressed and its $m_{HH}$ distribution broadens toward higher mass. At $\kappa_3=0.4$ the threshold enhancement survives and $m_{HH}$ keeps a sharp peak near $2m_H$. The background $m_{HH}$ is dominated by the $W^+W^-\nu\bar\nu$ and $ZZ\nu\bar\nu$ processes, which peak at low mass and overlap with the signal, so $m_{HH}$ mainly separates the $\kappa_3$ hypotheses rather than signal from background.
Because $H_0$ carries the information that separates the signal from the background while $m_{HH}$ carries the $\kappa_3$ information, the two are exploited respectively by the separately trained classifiers, and this division of roles will reappear at the feature level in the interpretability analysis of Sec.~\ref{sec:interpretability}.

\subsection{Boosted region}\label{sec:ml_boosted}

Compared to the resolved region, the model design in the boosted region is much simpler, since the candidate signal events have no pairing problem. Each candidate of the large-radius jets already contains a full $H\to b\bar b$ decay. A total of 16 high-level features, covering the leading and subleading large-radius jets and the event-level kinematics, and 160 low-level particle cloud features are taken into account.

\begin{figure}[t]
    \centering
    \includegraphics[width=\linewidth]{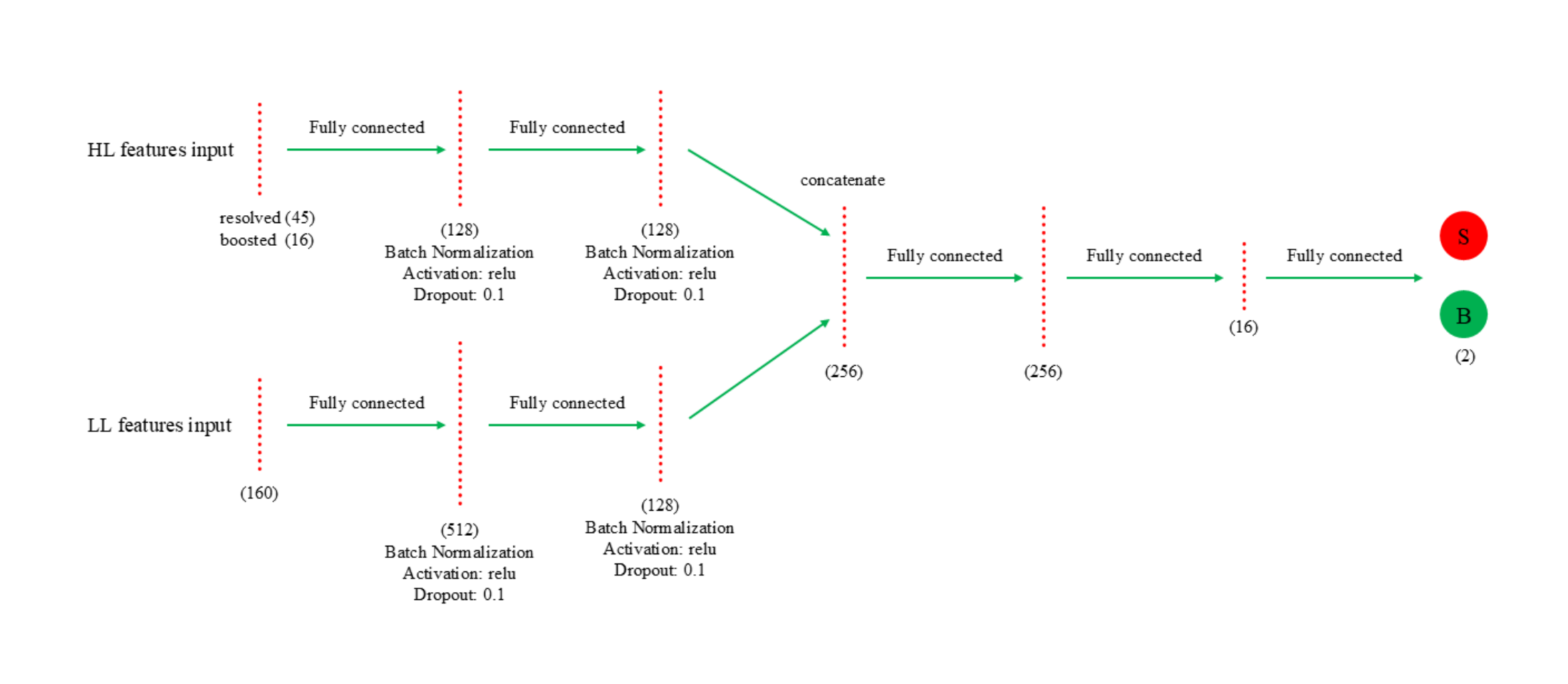}
    \caption{Architecture of the $\mathcal{D}_{\rm HH}$ network in the boosted region, which separates $HH$ production from the non-$HH$ background.}
    \label{fig:network_boosted}
\end{figure}

Fig.~\ref{fig:network_boosted} shows the $\mathcal{D}_{\rm HH}$ architecture in the boosted region. The model consists of one branch for the high-level features and one for the low-level features. Each branch is passed through two dense layers before the two are concatenated, and a few further dense layers lead to a sigmoid output that discriminates $HH$ production from the background.
The boosted $\mathcal{D}_{\rm HH}$ is implemented in \texttt{Keras}~\cite{chollet2015keras} as well, whereas the boosted $\mathcal{D}_{\kappa_3}$ uses the XGBoost algorithm~\cite{Chen:2016btl}. The classifier is trained with a binary logistic loss and uses 100 estimators, a learning rate of $0.1$, a maximum tree depth of $6$, and a pruning parameter $\gamma=6$ to mitigate overfitting and control the tree complexity.
On the held-out test sample the boosted $\mathcal{D}_{\rm HH}$ classifier reaches an area under the ROC curve of $0.99$ ($0.98$) at $\sqrt{s}=3$ ($10$)\,TeV. The statistical analysis based on $\mathcal{D}_{\rm HH}$ and $\mathcal{D}_{\kappa_3}$ is presented in Sec.~\ref{sec:statistics}.

\begin{figure}[t]
    \centering
    \includegraphics[width=\linewidth]{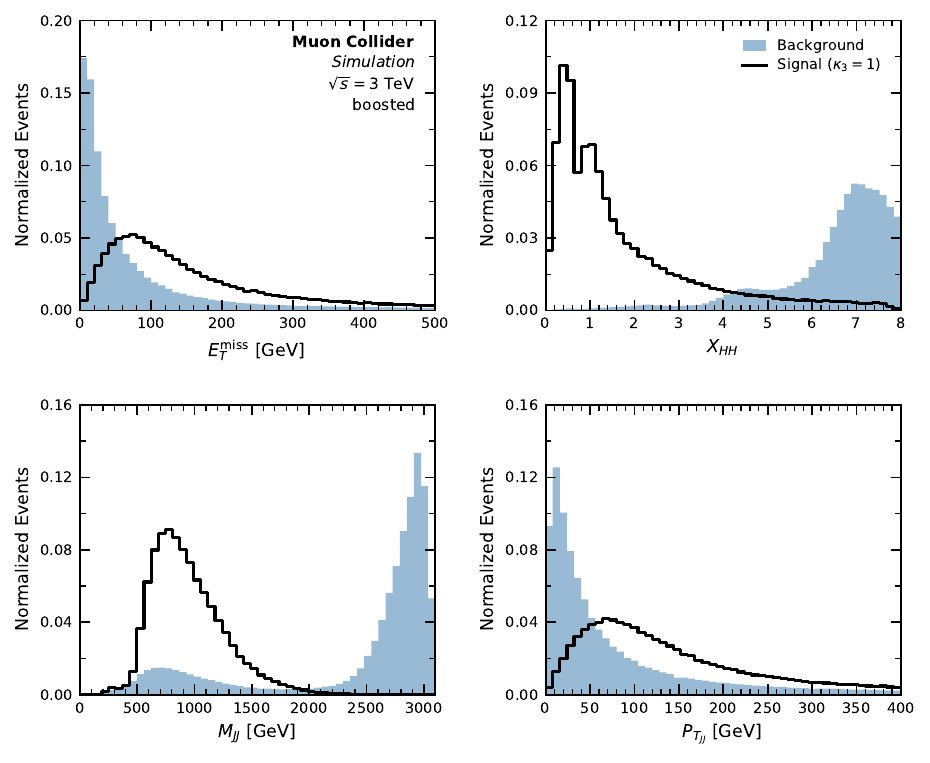}
    \caption{Representative boosted region observables at $\sqrt{s}=3$\,TeV after the pre-selection: $E_T^{\rm miss}$, $X_{HH}$ of Eq.~(\ref{eq:XHH}) evaluated with the fat jet masses, $M_{JJ}$, and $P_{T_{JJ}}$. The black line shows the signal ($\kappa_3=1$) and the filled histogram the total background. The individual processes are weighted by their cross sections and each distribution is normalized to unity.}
    \label{fig:kinematics_boosted}
\end{figure}

The input features are grouped as follows:
\begin{itemize}
\item \textbf{Fat jets}: for each of the two leading large-radius jets, the mass $M(J)$, the transverse momentum $p_T(J)$, and the $b$-tag ${\rm BTag}(J)$, with the same $70\%$ $b$-tagging efficiency as in the resolved region.
\item \textbf{Global features}: the missing transverse energy $E_T^{\rm miss}$, the mass window variable $X_{HH}$ of Eq.~(\ref{eq:XHH}) evaluated with the fat jet masses $M(J_1)$ and $M(J_2)$ and with $M_1=124$\,GeV and $M_2=115$\,GeV, the invariant mass $M_{JJ}$ and the combined transverse momentum $P_{T_{JJ}}$ of the two fat jets, $\Delta\eta_{JJ} = |\eta(J_1)-\eta(J_2)|$, and the five TDA descriptors $H_0$, $H_1$, $S_0$, $S_1$, $LB_1$ (Sec.~\ref{sec:tda}).
\item \textbf{Particle cloud}: the low-level features of the two leading fat jets. In each fat jet the particles are sorted by transverse momentum and the leading 20 are selected. For each particle four quantities are included, $p_T$, $\eta$, $\phi$, and a particle type category.
The particle type category is the detected particle type from the different calorimeter layers, with the same five categories and labels ($1$--$5$) as in the resolved region. Fat jets with fewer than 20 particles are zero-padded, giving $2\times20\times4=160$ low-level inputs per event. Unlike the resolved particle cloud stream, these inputs enter the network as a fixed-length vector.
\end{itemize}

Fig.~\ref{fig:kinematics_boosted} shows four of the global features at $\sqrt{s}=3$\,TeV after the pre-selection. The signal accumulates at small $X_{HH}$, since both fat-jet masses lie close to the Higgs candidate peak positions, and at large $E_T^{\rm miss}$, carried away by the recoiling neutrinos of the VBF topology, whereas the dominant two-body backgrounds populate the small missing energy region. The remaining two panels separate the two topologies just as sharply. The signal $M_{JJ}$ peaks near $800$\,GeV, close to the boosted threshold of $p_T(H)> 200$\,GeV for both Higgs candidates, because the pair recoils against the forward neutrinos, whereas the two-body backgrounds put the full collision energy into the two jets and pile up near $\sqrt{s}$. The same recoil gives the signal a broad $P_{T_{JJ}}$ distribution peaking near $70$\,GeV, while the back-to-back backgrounds balance in the transverse plane and accumulate at $P_{T_{JJ}}$ close to zero.
These separations will emerge as the leading features for the importance of the boosted classifier in Sec.~\ref{sec:interpretability}.

\subsection{Classifier scores}\label{sec:scores}

\begin{figure}[t]
    \centering
    \includegraphics[width=\textwidth]{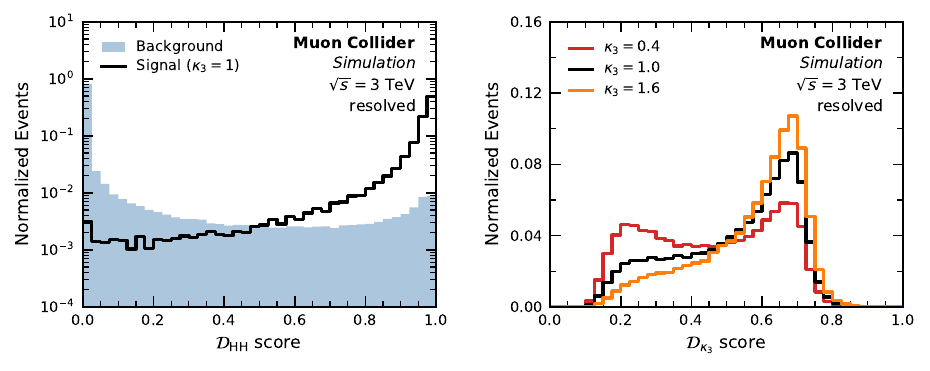}
    \caption{Output scores of the two classifiers at $\sqrt{s}=3$\,TeV in the resolved region. The left panel shows the $\mathcal{D}_{\rm HH}$ score for the $HH\to b\bar{b}b\bar{b}$ signal at $\kappa_3=1$ (black) and the inclusive background (filled blue). The right panel shows the $\mathcal{D}_{\kappa_3}$ score for three signal hypotheses, $\kappa_3=0.4$ (red), $\kappa_3=1.0$ (black), and $\kappa_3=1.6$ (orange).}
    \label{fig:scores_resolved}
\end{figure}

Fig.~\ref{fig:scores_resolved} shows the two classifier outputs at $\sqrt{s}=3$\,TeV in the resolved region.
In the left panel the network assigns a $\mathcal{D}_{\rm HH}$ score close to one to events that it identifies as signal and close to zero to those it identifies as background, so the signal peaks near one and the background near zero. The small background that survives at high $\mathcal{D}_{\rm HH}$ score is almost entirely $Hq\bar{q}\nu\bar\nu$, because this process shares the VBF topology of the signal and contains a genuine $H\to b\bar{b}$ decay. In the right panel the $\mathcal{D}_{\kappa_3}$ scores pile up near the middle, because for a single event the network cannot decide with confidence whether it corresponds to $\kappa_3=0.4$ or $\kappa_3=1.6$. Even so, the $\kappa_3=0.4$ distribution is shifted slightly to the left and the $\kappa_3=1.6$ distribution slightly to the right, and this residual separation is what $\mathcal{D}_{\kappa_3}$ contributes to the fit, and Sec.~\ref{sec:statistics} quantifies how much it is worth.

\begin{figure}[t]
    \centering
    \includegraphics[width=\linewidth]{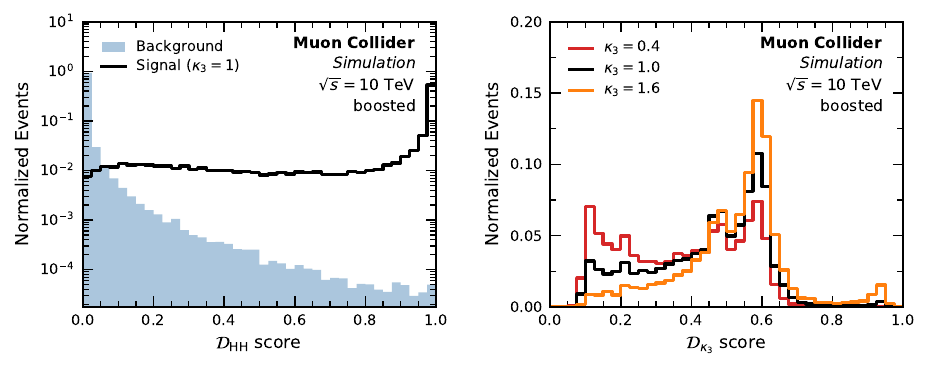}
    \caption{Output scores of the two classifiers at $\sqrt{s}=10$\,TeV in the boosted region. The left panel shows the $\mathcal{D}_{\rm HH}$ score for the $HH$ signal at $\kappa_3=1$ (black) and the inclusive background (filled blue). The right panel shows the $\mathcal{D}_{\kappa_3}$ score for three signal hypotheses, $\kappa_3=0.4$ (red), $\kappa_3=1.0$ (black), and $\kappa_3=1.6$ (orange). In the boosted channel both classifiers are used only at $\sqrt{s}=10$\,TeV (Sec.~\ref{sec:statistics}). The scores are therefore shown at that energy.}
    \label{fig:scores_boosted}
\end{figure}

Fig.~\ref{fig:scores_boosted} shows the output scores of the two classifiers at $\sqrt{s}=10$\,TeV in the boosted region. In the left panel the classifier pushes background events toward zero and signal events toward one. Most of the background indeed accumulates near zero, although some $t\bar t$ and $HZ\nu\bar\nu$ events remain close to one. The right panel presents the $\mathcal{D}_{\kappa_3}$ scores for $\kappa_3=0.4$, $1.0$, and $1.6$. The separation is weaker than for $\mathcal{D}_{\rm HH}$, since the three hypotheses share very similar kinematics, but the residual differences still improve the subsequent likelihood analysis.

\subsection{Classifier interpretability}\label{sec:interpretability}

\begin{figure}[t]
    \centering
    \includegraphics[width=\textwidth]{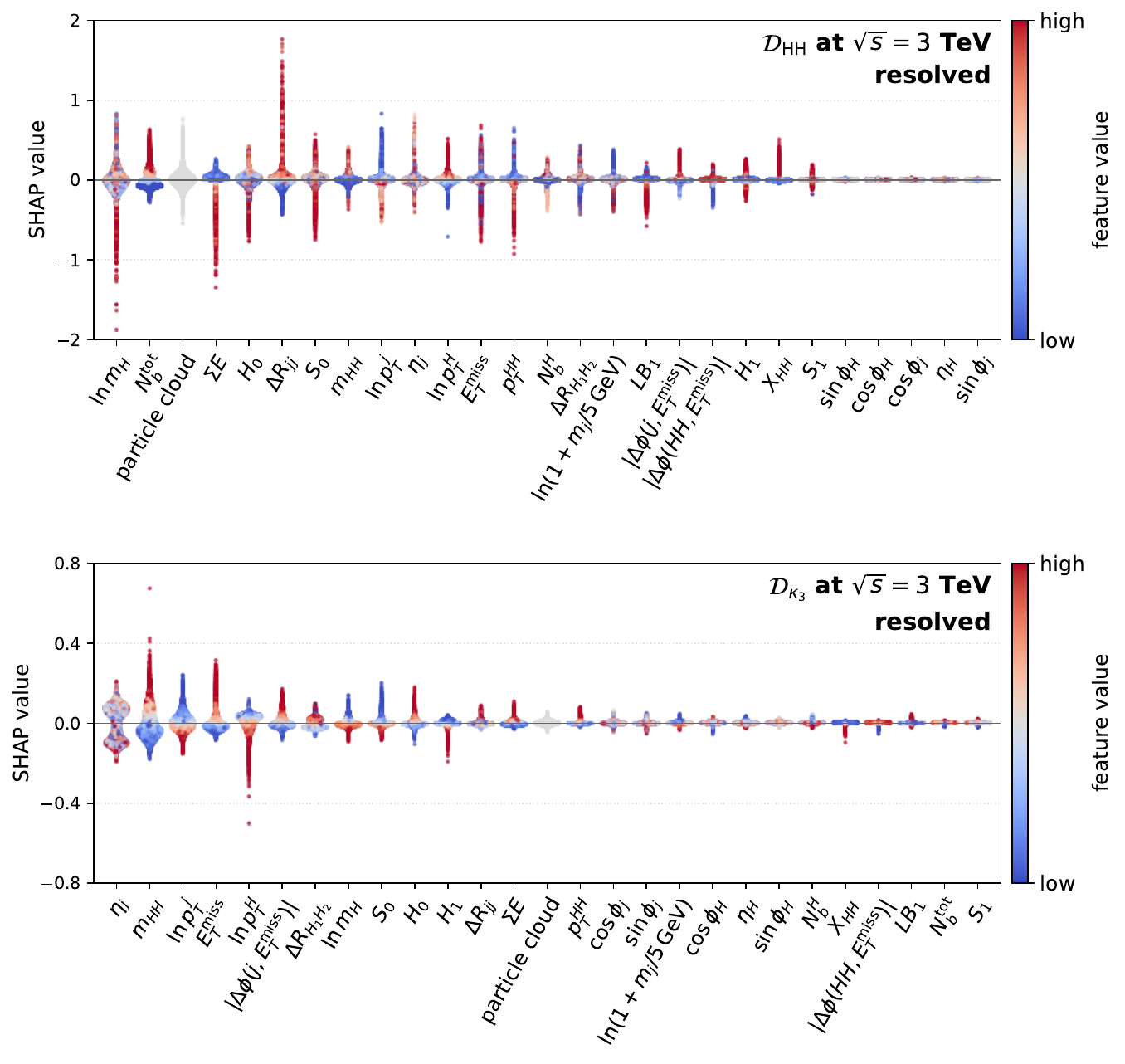}
    \caption{SHAP values of the input features of $\mathcal{D}_{\rm HH}$ (upper panel) and $\mathcal{D}_{\kappa_3}$ (lower panel) in the resolved region at $\sqrt{s}=3$\,TeV, ordered by decreasing mean absolute SHAP value.}
    \label{fig:shap_resolved}
\end{figure}

We examine which input features drive the classifiers in the two regions. To make their behavior interpretable, we compute SHAP (SHapley Additive exPlanations) values~\cite{Lundberg:2017une} for their input features, using the deep-model explainers of the SHAP library evaluated on the validation samples. 
Fig.~\ref{fig:shap_resolved} shows the result for $\mathcal{D}_{\rm HH}$ and $\mathcal{D}_{\kappa_3}$ in the resolved region at $\sqrt{s}=3$\,TeV.

Each column corresponds to one input feature and each point to one simulated event. The vertical position is the signed contribution of that feature to the network score for that event, positive values pushing it toward the signal side of $\mathcal{D}_{\rm HH}$ or toward the high-$\kappa_3$ side of $\mathcal{D}_{\kappa_3}$, and the color encodes the value of the feature itself, blue for low and red for high. Per-jet, per-candidate, and per-particle attributions are summed within each event, so the low-level stream appears as the single column labeled ``particle cloud'', whose points are drawn at a fixed zero feature value, since the summed contributions come from constituents with different feature values that no single color can represent. Within each panel the columns are ordered by the mean absolute SHAP value, independently for the two networks, with the most important feature on the left.

The $b$-tag multiplicity $N_b^{\rm tot}$, second in the upper panel, illustrates how such a distribution is read. Its points separate cleanly by color, which means that the feature acts monotonically. Events with no $b$-tagged jet, $43\%$ of the sample, appear in blue below the axis with a mean contribution of $-0.09$, while events with three or four tags, $15\%$ of the sample, appear in red above it with $+0.16$ and $+0.20$. The vertical spread within one color measures how much the same feature value can matter in different events. The blue group is narrow, with an rms of $0.03$, because an event without $b$-tagged jets looks like background whatever else it contains, whereas the red group is more than twice as broad, $0.07$, because a high $b$-tag multiplicity only helps when the rest of the event is signal-like as well. Averaging the absolute contribution over all events gives $0.08$ for this feature, which is what places it second in the ordering.

The two rankings confirm the design logic of the two-network strategy at the feature level. $\mathcal{D}_{\rm HH}$ is driven by the quantities that identify the $HH\to b\bar bb\bar b$ final state, namely the Higgs candidate mass, the $b$-tag multiplicity, the particle cloud, the scalar energy sum, and the TDA descriptor $H_0$, with the candidate mass acting non-monotonically, favoring the signal only near $m_H\simeq125$\,GeV. In contrast, $\mathcal{D}_{\kappa_3}$ is led by the production kinematics, the jet pseudorapidities and $m_{HH}$ foremost among them. Masses near the $2m_H$ threshold push the score toward the low-$\kappa_3$ hypothesis, where the threshold enhancement of the trilinear amplitude survives, and high masses toward the high-$\kappa_3$ hypothesis, where the destructive interference suppresses the threshold region. The $b$-tag multiplicity, second for $\mathcal{D}_{\rm HH}$, is among the least important features of $\mathcal{D}_{\kappa_3}$, since all $\kappa_3$ hypotheses share the same final state. This is the feature-level counterpart of the complementarity between $H_0$ and $m_{HH}$ anticipated in Fig.~\ref{fig:kinematics_resolved}.

\begin{figure}[t]
    \centering
    \includegraphics[width=\linewidth]{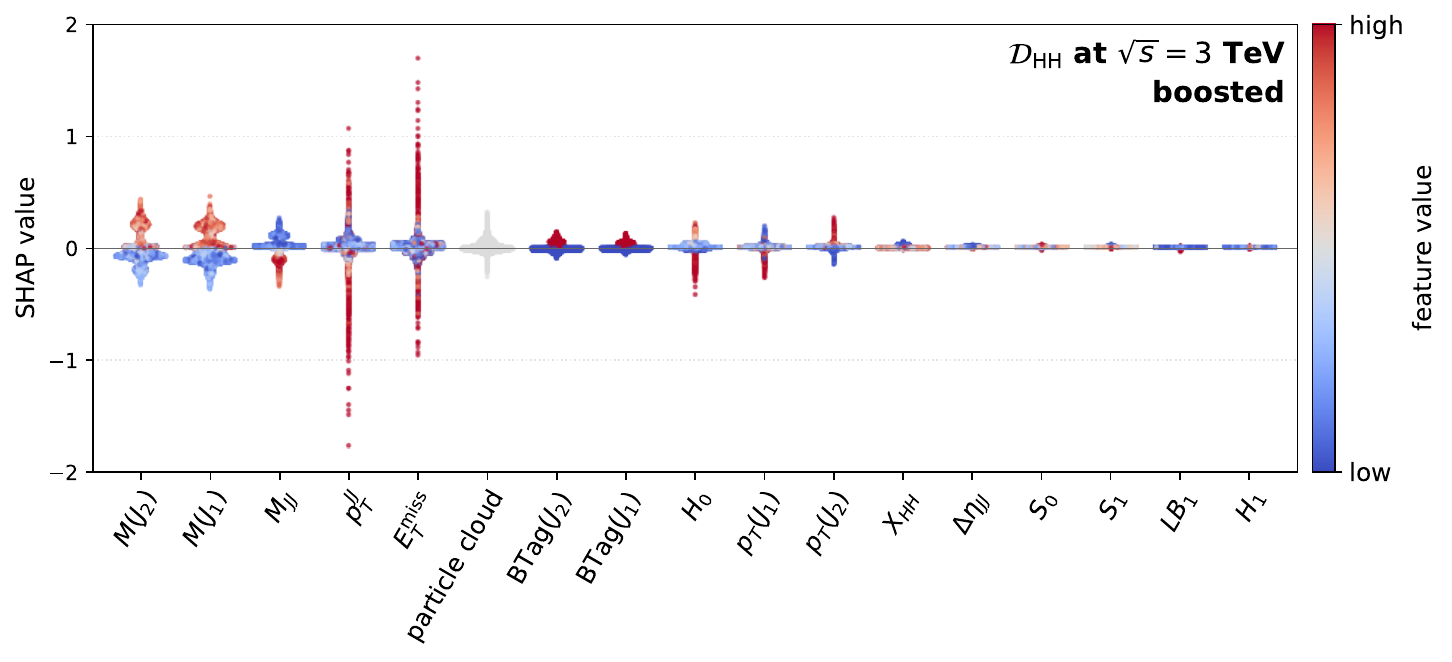}
    \caption{SHAP values of the input features of $\mathcal{D}_{\rm HH}$ in the boosted region at $\sqrt{s}=3$\,TeV, in the same style as Fig.~\ref{fig:shap_resolved}. A few extreme points beyond the displayed $|{\rm SHAP}|\le2$ range are clipped at the frame.}
    \label{fig:shap_boosted}
\end{figure}

The boosted region is examined in the same way. As Fig.~\ref{fig:shap_boosted} shows for the boosted $\mathcal{D}_{\rm HH}$, the two fat jet masses $M(J_2)$ and $M(J_1)$ are the leading features, as expected from the clear separation in Fig.~\ref{fig:preselection_boosted}, and both act in the same direction, favoring the signal for masses close to $m_H$. They are followed by the global kinematic variables $M_{JJ}$, $P_{T_{JJ}}$, and $E_T^{\rm miss}$, echoing the separations of Fig.~\ref{fig:kinematics_boosted} with directions dictated by the VBF topology, low $M_{JJ}$ and large $E_T^{\rm miss}$ favoring the signal, whose jet pair recoils against forward neutrinos that are absent in the $s$-channel backgrounds. In the summed particle cloud column, the dominant low-level feature is the particle type of the few most energetic constituents. Fat jets whose leading constituents include a lepton are preferentially classified as Higgs candidates, reflecting the well-known lepton-in-jet signature of $b$-hadron decays~\cite{ATLAS:2018kot}, and we verified at the generator level that these leptons originate almost exclusively from the $b$-decay chain.

The two regions agree on what matters most. The mass of the Higgs candidate, reconstructed from a jet pair in the resolved region and from a single large-radius jet in the boosted one, leads both rankings, with the $b$-tag information within the first few ranks, the expected hierarchy for a $b\bar bb\bar b$ final state. They differ in what they use next, the resolved classifier drawing on the event-level topology through the scalar energy sum, the particle cloud, and the descriptor $H_0$, and the boosted one on the far more constrained kinematics of the two-jet system. Among the five TDA descriptors, $H_0$ is the one that matters in both regions, ranking fifth for the resolved $\mathcal{D}_{\rm HH}$ and sitting level with the $b$-tag features in the boosted region, joined by $S_0$ in the resolved region, while $H_1$, $S_1$, and $LB_1$ stay near the bottom of every ranking. Built from the angular positions of the event constituents alone, without jet clustering, they describe the event in a way that no combination of the reconstructed jets reproduces, and including them tightens the resolved $68\%$ interval reported in Sec.~\ref{sec:results_k3}.

\subsection{Statistical analysis}\label{sec:statistics}

After the pre-selection, each event is processed by the two classifiers $\mathcal{D}_{\rm HH}$ and $\mathcal{D}_{\kappa_3}$, and $\kappa_3$ is extracted from a two-dimensional binned maximum likelihood fit to the distribution of their scores (Case~2 of Sec.~\ref{sec:introduction}, the discrimination between $\kappa_3$ hypotheses).
In the resolved region the $(\mathcal{D}_{\rm HH},\,\mathcal{D}_{\kappa_3})$ plane is divided into $10\times10$ bins, uniform in $\mathcal{D}_{\rm HH}$ and quantile-based in $\mathcal{D}_{\kappa_3}$, with the quantile edges determined from the background and SM-signal events. The background enters the same plane with a $\kappa_3$-independent template, so the response of $\mathcal{D}_{\kappa_3}$ to background events, on which it is never trained, can dilute but not bias the extracted coupling. The gain from the second network can be read off by discarding it. Binning in $\mathcal{D}_{\rm HH}$ alone widens the $68\%$ interval by $31\%$ and $33\%$ in the resolved region at $3$ and $10$\,TeV, and by $31\%$ in the boosted region at $10$\,TeV. The combined interval widens by $32\%$ at $10$\,TeV and by $22\%$ at $3$\,TeV, the smaller shift at $3$\,TeV following from the boosted channel there already using $\mathcal{D}_{\rm HH}$ alone. The $\kappa_3$ information carried by $\mathcal{D}_{\kappa_3}$ is therefore worth far more than the modest separation of its score distributions would suggest, because it is orthogonal to the signal-background axis rather than a sharper version of it.

For each $\kappa_3$ on the production grid (Sec.~\ref{sec:signal_background}) the per-bin yields of the nominal Monte Carlo simulation give the predicted spectrum.
The likelihood is the product of Poisson probabilities over the bins of that plane~\cite{ParticleDataGroup:2024cfk},
\begin{equation}                                              
\mathcal{L}(\kappa_3) = \prod_i \frac{[\mu_i(\kappa_3)]^{n_i} e^{-\mu_i(\kappa_3)}}{n_i!}, \qquad \mu_i(\kappa_3) = s_i(\kappa_3) + b_i, \label{eq:likelihood}
\end{equation}                                                
where $\mu_i$ is the expected number of events in bin $i$, given by the signal yield $s_i(\kappa_3)$ at that value of the coupling plus the background yield $b_i$, both taken from the nominal Monte Carlo simulation at the target luminosity, and $n_i$ is the number of events observed in the same bin. Since no data exist, the $n_i$ are the Asimov dataset~\cite{Cowan:2010js} at the SM point, $n_i = \mu_i(\kappa_3{=}1)$, built from a statistically independent $\kappa_3=1$ sample so that the likelihood is never evaluated against the same simulated events that build the templates. The coupling is probed with the test statistic
\begin{equation}                                              
-\Delta\ln L(\kappa_3) = \ln L(\kappa_3{=}1) - \ln L(\kappa_3).
\label{eq:dll}  
\end{equation}                
These values are fitted with a fourth-order polynomial. The best-fit $\hat\kappa_3$ is the position of its minimum, and the $68\%$ and $95\%$ confidence intervals are the connected ranges around $\hat\kappa_3$ in which the profile stays below $0.5$ and $1.92$ (single-parameter convention~\cite{ParticleDataGroup:2024cfk}).
The range of scan points entering each fit is stated together with the corresponding figure in Sec.~\ref{sec:results}. The Monte Carlo statistical uncertainty of the templates is suppressed by the enlarged near-minimum samples of Sec.~\ref{sec:signal_background}.

For the observability of $HH$ production above the SM background (Case~1 of Sec.~\ref{sec:introduction}), we use the same binned templates and the same test statistic, with the reference hypothesis of Eq.~(\ref{eq:dll}) replaced by the background-only one. It is evaluated, for each $\kappa_3$ on the scan grid, between that hypothesis and the sum of the background and $HH$ production. The corresponding expected significance for rejecting the background-only hypothesis is $Z=\sqrt{2\,(-\Delta\ln L)}$.
In the signal strength language of the LHC combination~\cite{CMS:2026nuu}, this is the expected significance with which the background-only hypothesis $\mu_{HH}=0$ is rejected if $HH$ production with the given $\kappa_3$ is present, where the signal strength $\mu_{HH}$ denotes the $HH$ rate in units of its predicted value.

The boosted channel at $\sqrt{s}=3$\,TeV requires a modified treatment. Table~\ref{tab:preselection_boosted} shows that fewer than ten signal events survive the pre-selection at the baseline luminosity, so the two-dimensional templates cannot be populated reliably and the resulting likelihood would overstate the significance. 
In this channel we therefore use the $\mathcal{D}_{\rm HH}$ classifier alone, with a score cut at $0.95$ that removes most of the background while preserving the signal. The expected significance increases monotonically with the cut value, but a cut tighter than $0.95$ would leave fewer than three expected signal events, so we fix the working point there. The same cut defines the boosted Case~1 observability.
The $\kappa_3$ dependence of the yield in this single-bin selection likewise provides the boosted $-\Delta\ln L(\kappa_3)$ at $3$\,TeV. At $10$\,TeV, where the signal yield is sufficient, the two-dimensional likelihood of Eq.~(\ref{eq:likelihood}) is used instead, with its own binning of the $(\mathcal{D}_{\rm HH},\,\mathcal{D}_{\kappa_3})$ plane.
With only a few expected events, the asymptotic conversion between $-\Delta\ln L$ and significance is approximate at $3$\,TeV in this channel, which is a further reason why the boosted region enters the $3$\,TeV results mainly as a supporting channel.

The sensitivities quoted in Sec.~\ref{sec:results} are statistical-only Asimov projections. Systematic uncertainties, such as the $b$-tagging efficiency, the background normalizations, and the luminosity, are not included, and the beam-induced background is not overlaid on the simulated events, its mitigation being assumed to be achievable by future detector and reconstruction improvements, as in the $HH$ analysis of Ref.~\cite{Andreetto:2024rra}.

%--------------------------------------------------------------------------------
\section{Results}\label{sec:results}

\subsection{Observability of Higgs boson pair production}\label{sec:observability}

We first address Case~1 of Sec.~\ref{sec:introduction}, the observability of $HH$ production above the SM background, following the procedure of Sec.~\ref{sec:statistics}.

\begin{figure}[t]
    \centering
    \includegraphics[width=\textwidth]{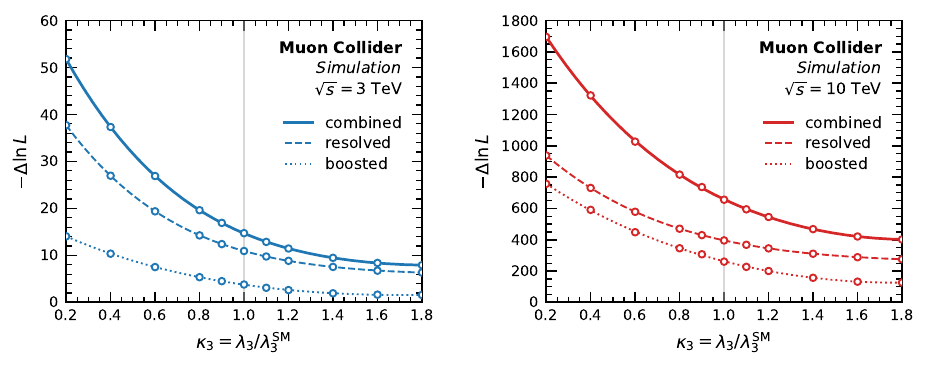}
    \caption{Expected $-\Delta\ln L$ for rejecting the background-only hypothesis in the presence of $HH$ production with trilinear coupling $\kappa_3$, at $\sqrt{s}=3$\,TeV (left) and $10$\,TeV (right). Each panel shows the three channels, boosted only (dotted), resolved only (dashed), and combined (solid). Open markers are the scanned $\kappa_3$ points, the vertical line marks the SM point $\kappa_3=1$, and the expected significance is $Z=\sqrt{2\,(-\Delta\ln L)}$.}
    \label{fig:dll_observability}
\end{figure}

Fig.~\ref{fig:dll_observability} shows the result at both energies for the resolved (dashed), boosted (dotted), and combined (solid) channels, where the last one is obtained by adding the two profiles as in Sec.~\ref{sec:results_combined}. At the SM point the resolved channel alone reaches an expected significance of $4.7\sigma$ at $\sqrt{s}=3$\,TeV and $28\sigma$ at $10$\,TeV, and the boosted channel alone reaches $2.8\sigma$ and $23\sigma$, respectively. Combining the two channels, the significance reaches $5.4\sigma$ at $3$\,TeV and $36\sigma$ at $10$\,TeV, so the SM $HH$ signal is observable already at $3$\,TeV and overwhelmingly so at $10$\,TeV. The observability decreases monotonically with $\kappa_3$, tracking the total cross section toward its interference minimum near $\kappa_3\simeq1.7$. Even there the combined significance remains $4.0\sigma$ at $3$\,TeV and above $28\sigma$ at $10$\,TeV. The $\kappa_3$ measurement presented below therefore proceeds on top of an observed signal at both energies.

\subsection{Sensitivity to the trilinear coupling}\label{sec:results_k3}

We now turn to Case~2 of Sec.~\ref{sec:introduction}, the discrimination of $\kappa_3\neq1$ from the SM point, quantified by the two-dimensional likelihood fit of Sec.~\ref{sec:statistics}. The resolved and boosted regions are presented first, followed by their combination.

\subsubsection{Resolved region}\label{sec:results_resolved}

\begin{figure}[t]
\centering
    \includegraphics[width=\textwidth]{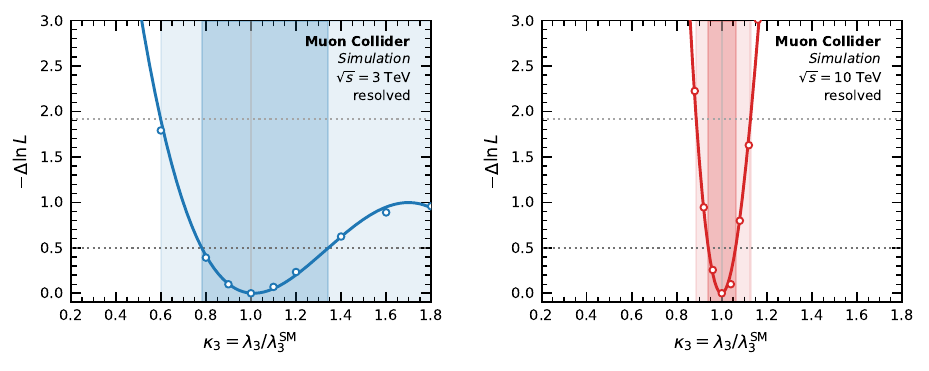}
    \caption{Resolved $-\Delta\ln L(\kappa_3)$ profile at $\sqrt{s}=3$\,TeV (left) and $10$\,TeV (right). Open markers are the scanned $\kappa_3$ points and the solid curve is a fourth-order polynomial fit through them. At $10$\,TeV the fit uses only the finer points with $\kappa_3\in[0.8,1.2]$. The darker (lighter) bands mark the $68\%$ ($95\%$) confidence intervals, bounded by the horizontal dotted lines at $-\Delta\ln L=0.5$ ($1.92$).}
    \label{fig:dll_resolved}
\end{figure}

Fig.~\ref{fig:dll_resolved} shows the resolved $-\Delta\ln L(\kappa_3)$ profile at the two center-of-mass energies.
These profiles are fourth-order polynomial fits to the scanned $\kappa_3$ points. The scan uses the full grid of Eq.~(\ref{eq:k3grid}) at both $\sqrt{s}=3$ and $10$\,TeV, and the finer points with $\kappa_3\in[0.8,1.2]$ of Eq.~(\ref{eq:k3grid_10tev}) at $10$\,TeV, which scan with higher precision at the narrow trough $\kappa_3=1$. There the profile rises so steeply away from $\kappa_3=1$ that points far from the minimum would otherwise dominate the fit and distort the shape of the trough, which eventually sets the $68\%$ and $95\%$ intervals.
The likelihood profile has a clearly defined minimum at both energies and exhibits the asymmetry anticipated in Sec.~\ref{sec:hh_production}.
The right flank ($\kappa_3>1$) is shallower than the left one ($\kappa_3<1$) in Fig.~\ref{fig:dll_resolved}, because the $\kappa_3$-dependent part of the cross section turns over near $\kappa_3^{\rm min}\!\simeq\!1.7$ and the rate variation slows down for $\kappa_3>1$.

At $\sqrt{s}=3$\,TeV with $\mathcal{L}=1$\,ab$^{-1}$, the resolved $68\%$ and $95\%$ confidence intervals are
\begin{equation}                                              
0.78 \;<\; \kappa_3 \;<\; 1.34 \quad (68\%~{\rm CL}),         
\qquad                                                        
0.60 \;<\; \kappa_3 \;<\; \text{---} \quad (95\%~{\rm CL}),   
\label{eq:cl_resolved_3tev}                                   
\end{equation}     
where the upper $95\%$ edge is not closed within the scanned $\kappa_3 \in[0.2,1.8]$ range. 

At $\sqrt{s}=10$\,TeV with $\mathcal{L}=10$\,ab$^{-1}$, the resolved confidence intervals tighten substantially to
\begin{equation}
0.94 \;<\; \kappa_3 \;<\; 1.06 \quad (68\%~{\rm CL}), 
\qquad
0.89 \;<\; \kappa_3 \;<\; 1.13 \quad (95\%~{\rm CL}), 
\label{eq:cl_resolved_10tev}
\end{equation}
nearly a factor of five narrower than at $3$\,TeV. The gain is the combined effect of a tenfold increase in integrated luminosity and a $\sim 4$-fold increase in the production cross section at $\kappa_3=1$, partially offset by the reduced resolved channel acceptance noted in Sec.~\ref{sec:signal_background}.

\subsubsection{Boosted region}\label{sec:results_boosted}

\begin{figure}[t]
    \centering
    \includegraphics[width=\linewidth]{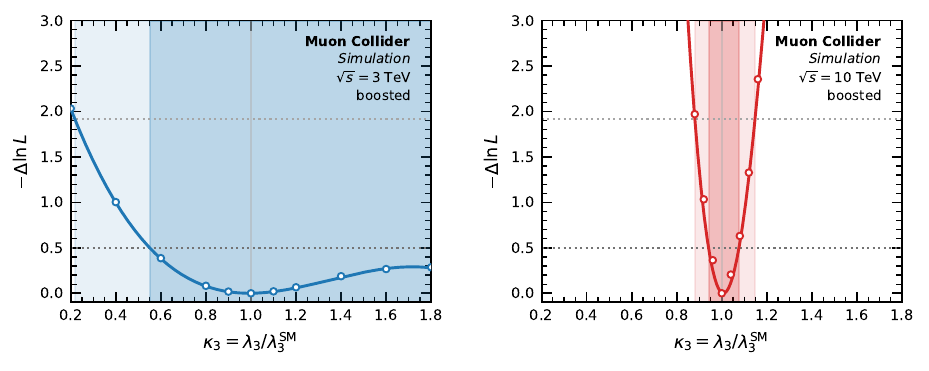}
    \caption{Boosted $-\Delta\ln L(\kappa_3)$ profile at $\sqrt{s}=3$\,TeV (left) and $10$\,TeV (right), with the same conventions as Fig.~\ref{fig:dll_resolved}. At $10$\,TeV the scanned points outside the displayed well lie far above the vertical range.}
    \label{fig:dll_boosted}
\end{figure}

Fig.~\ref{fig:dll_boosted} shows the boosted $-\Delta\ln L(\kappa_3)$ profiles at $\sqrt{s}=3$ and $10$\,TeV, referenced to the $\kappa_3=1$ hypothesis and presented with the same conventions as the resolved profiles of Fig.~\ref{fig:dll_resolved}. The $68\%$ ($95\%$) confidence intervals on $\kappa_3$ are read off as the connected regions where the profile satisfies $-\Delta\ln L<0.5$ ($1.92$).

At $\sqrt{s}=3$\,TeV with $\mathcal{L}=1$\,ab$^{-1}$ the boosted region $68\%$ and $95\%$ confidence intervals are
\begin{equation}
    0.55 \;<\; \kappa_3  \;<\; \text{---} \quad (68\%~{\rm CL}),
    \qquad
    0.22 \;<\; \kappa_3  \;<\; \text{---} \quad (95\%~{\rm CL}),
\label{eq:cl_boosted_3tev}
\end{equation}
with no upper bounds within the scanned range.

As shown in Sec.~\ref{sec:sm_processes}, the cross section reaches its minimum near $\kappa_3\approx1.7$ and rises again beyond it. We therefore expect an upper bound to appear at larger $\kappa_3$, and in any case the upper edge is already bounded by the resolved region analysis.
At $\sqrt{s}=10$\,TeV with $\mathcal{L}=10$\,ab$^{-1}$ the boosted region $68\%$ and $95\%$ confidence intervals are shortened to
\begin{equation}
    0.94 \;<\; \kappa_3 \;<\; 1.08 \quad (68\%~{\rm CL}), 
    \qquad
    0.88 \;<\; \kappa_3 \;<\; 1.15 \quad (95\%~{\rm CL}), 
\label{eq:cl_boosted_10tev}
\end{equation}
respectively, far tighter than in the $3$\,TeV case.

The boosted result at $10$\,TeV far exceeds that at $3$\,TeV for three main reasons. 
First, a larger fraction of the Higgs bosons is boosted at $10$\,TeV, which increases the accessible cross section. Second, the $10$\,TeV design luminosity is ten times larger. Third, the event yield is large enough to apply the full two-dimensional likelihood analysis. This validates the expectation that the boosted region becomes more important as the center-of-mass energy increases.

\subsubsection{Combined results}\label{sec:results_combined}

\begin{figure}[t]
    \centering
    \includegraphics[width=\textwidth]{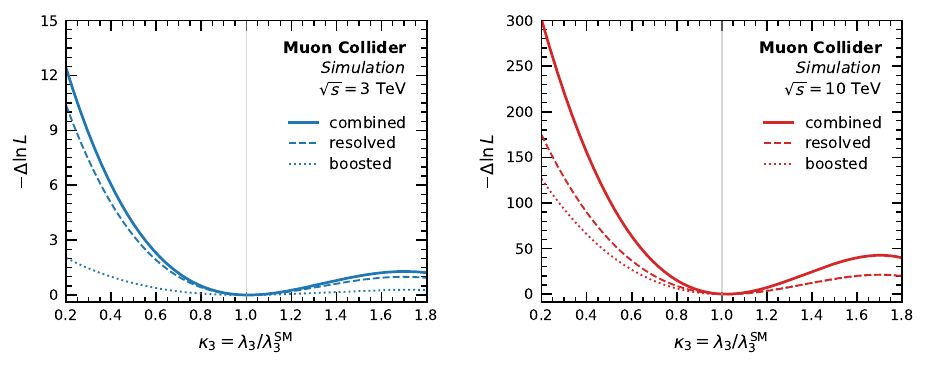}
    \caption{Combined resolved $\oplus$ boosted $-\Delta\ln L(\kappa_3)$ profile at $\sqrt{s}=3$\,TeV (left) and $10$\,TeV (right). Each panel shows the three channels, boosted only (dotted), resolved only (dashed), and combined (solid). Every curve is a fourth-order polynomial fit and the vertical line marks $\kappa_3=1$. Note the different vertical scales of the two panels.}
    \label{fig:dll_combined}
\end{figure}

The resolved and boosted analyses select largely non-overlapping kinematic regions.
At $\sqrt{s}=10$\,TeV, for instance, two thirds of the SM signal has both Higgs candidates below the $p_T(H)<200$\,GeV resolved threshold, while the boosted analysis captures the $12\%$ of events in which both candidates exceed it (events with one candidate on each side of the threshold enter neither analysis).
The two channels can therefore be treated as independent measurements and combined at the likelihood level by adding their $-\Delta\ln L(\kappa_3)$ profiles, both referred to the $\kappa_3=1$ Asimov hypothesis. Fig.~\ref{fig:dll_combined} shows the combined profile at the two energies.
The three curves shown there are fitted to the full scan grid at both energies, so that the figure displays the magnitude of $-\Delta\ln L$ over the whole range rather than only the well. The confidence intervals quoted below instead follow the fit ranges of Figs.~\ref{fig:dll_resolved} and \ref{fig:dll_boosted}.

At $\sqrt{s}=3$\,TeV with $\mathcal{L}=1$\,ab$^{-1}$, the combined $68\%$ and $95\%$ confidence intervals are
\begin{equation}
0.80 \;<\; \kappa_3 \;<\; 1.29 \quad (68\%~{\rm CL}), 
\qquad
0.63 \;<\; \kappa_3 \;<\; \text{---}\quad (95\%~{\rm CL}), 
\label{eq:cl_combined_3tev}
\end{equation}
where the upper $95\%$ edge is again pinned to the scan boundary. The combined interval is narrower than the resolved-only result ($0.78 < \kappa_3 < 1.34$), while the boosted channel alone does not close a $68\%$ upper bound at $3$\,TeV within the explored range.

At $\sqrt{s}=10$\,TeV with $\mathcal{L}=10$\,ab$^{-1}$, the combined intervals become
\begin{equation}
0.96 \;<\; \kappa_3 \;<\; 1.05 \quad (68\%~{\rm CL}), 
\qquad
0.92 \;<\; \kappa_3 \;<\; 1.10 \quad (95\%~{\rm CL}), 
\label{eq:cl_combined_10tev}
\end{equation}
constraining $\kappa_3$ to within approximately $\pm5\%$ at $68\%$ CL.
At both energies the combination improves on either channel alone, reflecting the kinematic complementarity of the resolved and boosted regions, and shows that a multi-TeV muon collider can constrain the Higgs trilinear self-coupling (and with it the shape of the electroweak vacuum) at the few-percent level.

Table~\ref{tab:cl_summary} collects the $68\%$ and $95\%$ CL intervals on $\kappa_3$ extracted from the three channels at the two energies. The $95\%$ CL at $\sqrt{s}=10$\,TeV is explicitly tabulated for all three channels. At $\sqrt{s}=3$\,TeV the $95\%$ CL upper boundary is not closed within the scanned $\kappa_3\in[0.2,1.8]$ window for any of the three channels, and the $3$\,TeV $95\%$ column is therefore omitted from the table. We expect the upper bound to lie above $\kappa_3\approx 1.8$, consistent with the upturn of $\sigma(\kappa_3)$ for $\kappa_3>1.7$ visible in Table~\ref{tab:preselection_signal_resolved}.
A dedicated resolved-channel scan up to $\kappa_3=3$ confirms this expectation. As $\sigma(\kappa_3)$ recovers beyond its minimum (Fig.~\ref{fig:xsec_kappa3}), the $3$\,TeV resolved $95\%$ interval closes at approximately $0.60<\kappa_3<2.5$, with a shallow, quasi-degenerate second likelihood minimum near $\kappa_3\approx2$ that stays above the $68\%$ threshold. The parametric $3$\,TeV studies of Refs.~\cite{MuonCollider:2022xlm,Accettura:2023ked} find the analogous minimum at $1.85<\kappa_3<1.94$, where it is deep enough to split their $68\%$ bound into two intervals. In our analysis it stays above the threshold and the bound remains connected.\footnote{%
The scan uses dedicated resolved channel samples generated at $\kappa_3 = 2.5$ and $3.0$ ($2.0$, $2.5$, and $3.0$ at $\sqrt{s}=10$\,TeV). The local minimum near $\kappa_3\approx2$ reaches $-\Delta\ln L\simeq0.9$. At $\sqrt{s}=10$\,TeV the resolved $-\Delta\ln L$ remains far above both thresholds throughout $1.8<\kappa_3\le 3$ and the quasi-degenerate second solution is absent.}

\begin{table}[t]
    \centering
    \setlength{\tabcolsep}{8pt}
    \begin{tabular}{lccc}
    \hline\hline
    Channel & $\sqrt{s}=3$\,TeV $68\%$ CL & $\sqrt{s}=10$\,TeV $68\%$ CL & $\sqrt{s}=10$\,TeV $95\%$ CL \\
    \hline
    Resolved             & $0.78 < \kappa_3 < 1.34$                 & $0.94 < \kappa_3 < 1.06$ & $0.89 < \kappa_3 < 1.13$ \\
    Boosted              & $0.55 < \kappa_3 < {}$\quad---           & $0.94 < \kappa_3 < 1.08$ & $0.88 < \kappa_3 < 1.15$ \\
    Combined             & $0.80 < \kappa_3 < 1.29$                 & $0.96 < \kappa_3 < 1.05$ & $0.92 < \kappa_3 < 1.10$ \\
    \hline
    Refs.~\cite{Andreetto:2024rra,Andreetto:2026flp}
                         & $0.81 < \kappa_3 < 1.44$    & $0.94 < \kappa_3 < 1.08$ & --- \\
    \hline\hline
    \end{tabular}\\[0.4em]
    \caption{Final $\kappa_3$ sensitivity. $68\%$ and $95\%$ CL intervals on $\kappa_3$ obtained at $\sqrt{s}=3$\,TeV ($\mathcal{L}=1$\,ab$^{-1}$) and $\sqrt{s}=10$\,TeV ($\mathcal{L}=10$\,ab$^{-1}$) from the resolved, the boosted, and the combined VBF $HH\!\to\!b\bar{b}b\bar{b}$ analyses. All intervals are read from the connected $-\Delta\ln L<0.5$ and $<1.92$ regions of a fourth-order polynomial fit. The $3$\,TeV $95\%$ CL column is omitted because the upper boundary is not closed within the nominal scan range $\kappa_3\in[0.2,1.8]$, and the extended scan is discussed in the text.
    The dash in the Boosted row marks a $68\%$ CL upper boundary that is not closed within the scanned $\kappa_3\in[0.2,1.8]$ range. The last row lists, for comparison, the $68\%$ CL intervals obtained at the same energies and integrated luminosities by Ref.~\cite{Andreetto:2024rra} ($3$\,TeV) and Ref.~\cite{Andreetto:2026flp} ($10$\,TeV), with the dash marking a value not reported.}
    \label{tab:cl_summary}
\end{table}

The profiles of Fig.~\ref{fig:dll_combined} also quantify how strongly larger departures from the SM point are disfavored. At $\sqrt{s}=10$\,TeV the combined profile crosses $-\Delta\ln L=12.5$ at $\kappa_3\simeq0.82$ and $1.27$, so any trilinear coupling outside this range would be excluded at more than $5\sigma$, and at the edge of the scanned range, $\kappa_3=0.2$, it reaches $-\Delta\ln L=301$, with the resolved and boosted channels contributing $175$ and $126$, respectively. At $\sqrt{s}=3$\,TeV the combined profile reaches $-\Delta\ln L=12.4$ at $\kappa_3=0.2$, corresponding to $5.0\sigma$.

The $68\%$ intervals are to be compared with $0.81<\kappa_3<1.44$ at $3$\,TeV~\cite{Andreetto:2024rra} and $0.94<\kappa_3<1.08$ at $10$\,TeV~\cite{Andreetto:2026flp}, obtained at the same energies and integrated luminosities. Subsequent studies assuming two interaction points~\cite{Accettura:2023ked}, each collecting $10$\,ab$^{-1}$, report $0.96<\kappa_3<1.06$ at $10$\,TeV~\cite{Andreetto:2026pyu,Andreetto:2026gvm}.
Ours, $0.80<\kappa_3<1.29$ and $0.96<\kappa_3<1.05$, are about $20\%$ narrower at $3$\,TeV and about $30\%$ narrower at $10$\,TeV, although the two analyses are not directly comparable. Ref.~\cite{Andreetto:2024rra} keeps two inclusive four-jet classes, both restricted to heavy quarks, $q_h=b,c$, light jets being taken as negligible once $b$ tagging is applied. Our samples cover all five quark flavors, so they retain the $W^+W^-$ and $t\bar t$ final states that such a requirement would remove, generated both with and without the accompanying $\nu\bar\nu$, and the boosted region adds two-body $b\bar b$ and light $q\bar q$ production. Our $b$-tagging efficiency is also lower, $70\%$ against $76\%$. The strength of the present analysis is clearest away from the SM point, in the steep profiles quoted above, which close the $2\sigma$ and $3\sigma$ intervals at $0.92<\kappa_3<1.10$ and $0.87<\kappa_3<1.15$.
Unlike Ref.~\cite{Andreetto:2024rra}, which suppresses the background with an explicit requirement of at least three $b$-tagged jets, we impose no $b$-tag multiplicity cut. The per-jet and total $b$-tag information is instead provided to the classifiers as input features, letting the networks exploit it directly.

%--------------------------------------------------------------------------------
\section{Conclusion}\label{sec:conclusion}

We have explored the sensitivity of multi-TeV muon colliders to the Higgs trilinear coupling $\kappa_3$ through $HH$ production via vector boson fusion, with both Higgs bosons decaying to $b\bar{b}$, at $\sqrt{s}=3$\,TeV and $10$\,TeV. The $b\bar{b}b\bar{b}$ final state is analyzed in two complementary regions defined by the Higgs transverse momentum. When $p_T(H)\lesssim 200$\,GeV the two $H\to b\bar{b}$ decays are resolved into four separate $b$-jets, whereas at higher $p_T$ each decay is collimated into a single large-radius jet and the event is treated in the boosted region.

Two networks are trained separately in each region on different objectives. $\mathcal{D}_{\rm HH}$ targets the separation of the signal from the background, and $\mathcal{D}_{\kappa_3}$ targets the shape variation of the signal across $\kappa_3$ hypotheses. In the resolved region a SPANet network first assigns the four jets to two Higgs candidates. Each classifier then takes four input streams built from the jets, their per-jet particle clouds, global event features including the TDA descriptors, and the reconstructed Higgs candidates. In the boosted region each Higgs decay is already contained in a single large-radius jet, so the pairing step is unnecessary. The boosted $\mathcal{D}_{\rm HH}$ takes two streams, the high-level event features and the low-level particle cloud features, while $\mathcal{D}_{\kappa_3}$ is a gradient-boosted decision tree on the high-level features. The coupling is then extracted from a binned maximum likelihood fit in the $(\mathcal{D}_{\rm HH},\,\mathcal{D}_{\kappa_3})$ plane, reduced to $\mathcal{D}_{\rm HH}$ alone in the statistics-limited $3$\,TeV boosted region.

The SM $HH$ signal itself is observable above the background-only hypothesis with an expected significance of $5.4\sigma$ at $\sqrt{s}=3$\,TeV and $36\sigma$ at $10$\,TeV for the combined channel.
Combining the resolved and boosted regions, the $68\%$ confidence intervals on $\kappa_3$ are $0.80<\kappa_3<1.29$ at $\sqrt{s}=3$\,TeV with $1$\,ab$^{-1}$ and $0.96<\kappa_3<1.05$ at $\sqrt{s}=10$\,TeV with $10$\,ab$^{-1}$, the latter broadening to $0.92<\kappa_3<1.10$ at $95\%$ confidence. The marked improvement from $3$ to $10$\,TeV follows from the larger integrated luminosity, the rising vector boson fusion cross section, and the growing weight of the boosted topology at higher energy. Note that the estimated sensitivity for $\kappa_3$ at HL-LHC is only $ 0.5 < \kappa_3 < 1.6$ at 68\% confidence level, which is not even as good as a 3 TeV muon collider with 1 ab$^{-1}$  luminosity.

These projections place the muon collider prominently in the landscape of future $\kappa_3$ determinations. At $e^+e^-$ Higgs factories operating below the $HH$ threshold, the self-coupling is accessible only indirectly, through its loop contributions to single-Higgs and $ZH$ observables, with a projected precision on $\kappa_3$ of $\mathcal{O}(30\%)$ at the FCC-ee, the exact value depending on the flavor assumptions of the fit~\cite{Maura:2025rcv}. Direct determinations from $HH$ production at a $100$\,TeV hadron collider with $30$\,ab$^{-1}$ are expected to reach $3$--$6\%$~\cite{Park:2020yps,Mangano:2020sao,Stapf:2023ndn}. The $10$\,TeV muon collider therefore approaches the ultimate hadron collider precision with a direct, tree-level measurement in vector boson fusion, and the three approaches probe the coupling through complementary production mechanisms and energy regimes.

These results show that a multi-TeV muon collider can constrain the Higgs self-coupling at the few-percent level, a precision that would directly probe the shape of the Higgs potential and the dynamics of electroweak symmetry breaking, and would decisively surpass the projected HL-LHC sensitivity~\cite{ATLAS:2022faz}.

%--------------------------------------------------------------------------------
\acknowledgments
The work of K.C., S.L., and C.W. is supported by the National Science \& Technology Council under grant no. NSTC 113-2112-M-007-041-MY3.
The work of J.S.L. was supported by the NRF grant funded by the Korea government (No.\ RS-2025-23523535).
Claude (Anthropic) was used to assist with debugging parts of SPANet and the classifiers for the resolved region described in Sec.~\ref{sec:ml_resolved}. Gemini (Google) was used to draft the LaTeX source of Fig.~\ref{fig:feynman_diagrams} and to format some tables. All suggestions from AI were carefully reviewed and validated by the authors.

%--------------------------------------------------------------------------------
\bibliographystyle{JHEP}
\bibliography{references}

\end{document}